\documentclass[12pt]{article}
\usepackage{amsmath,latexsym,amssymb,epsfig}
\newcommand{\beeq}{\begin{equation}}
\newcommand{\ene}{\end{equation}}
\newcommand{\bea}{\begin{eqnarray}}
\newcommand{\ena}{\end{eqnarray}}

\newcommand{\nb}{\nonumber}

\newcommand{\Bigbreak}{\par \ifdim\lastskip < \bigskipamount \removelastskip
\fi \penalty-300 \vskip 10mm plus 5mm minus 2mm}

\newcommand{\1}{{\bf 1}}
\newcommand{\e}{{\rm e}}

\newcommand{\F}{{\cal F}}

\newcommand{\A}{{\cal A}}

\newcommand{\Q}{{\cal Q}}
\newcommand{\T}{{\cal T}}

\newcommand{\R}{\mathbb{R}}

\newcommand{\der}{\partial }
\newcommand{\mis}{\frac{dk}{ 2 \pi }}

\newcommand{\sci}{\varphi }
\newcommand{\dsc}{\widetilde \varphi}
\newcommand{\scr}{\varphi_{{}_R}}
\newcommand{\scl}{\varphi_{{}_L}}
\newcommand{\scz}{\varphi_{{}_Z}}
\newcommand{\sczu}{\varphi_{{}_{Z_1}}}
\newcommand{\sczd}{\varphi_{{}_{Z_2}}}
\newcommand{\dlt}{\delta_{{}_{Z_1}{}_{Z_2}}}

\newcommand{\fz}{f_{{}_Z}}
\newcommand{\jz}{j_{{}_Z}}
\newcommand{\az}{\A_{{}_Z}}
\newcommand{\Qr}{Q_{{}_R}}
\newcommand{\Ql}{Q_{{}_L}}
\newcommand{\Qz}{Q_{{}_Z}}
\newcommand{\qr}{q_{{}_R}}
\newcommand{\ql}{q_{{}_L}}
\newcommand{\qz}{q_{{}_Z}}

\newcommand{\cqz}{{\Q}_{{}_Z}(f_{{}_Z})}
\newcommand{\cqzu}{{\Q}_{{}_{Z_1}}(f_{{}_{Z_1}})}
\newcommand{\cqzd}{{\Q}_{{}_{Z_2}}(f_{{}_{Z_2}})}
\newcommand{\spr}{\psi_{{}_R}}
\newcommand{\spl}{\psi_{{}_L}}
\newcommand{\ep}{{\e^{i|k|x^0-ikx^1}}}
\newcommand{\emi}{{\e^{-i|k|x^0+ikx^1}}}

\begin{document}

\thispagestyle{empty} 

\rightline {IFUP-TH 25/99}

\vskip 0.3 truecm
\centerline {\Large \bf Bosonization at Finite Temperature}
\bigskip 
\centerline {\Large \bf and Anyon Condensation}
\vskip 1 truecm
\centerline {\large \rm Antonio Liguori}
\medskip
\centerline {\it Dipartimento di Fisica dell'Universit\`a di Pisa,}
\centerline {\it Via Buonarroti 2, 56127 Pisa, Italy} 
\bigskip
\medskip
\centerline {\large \rm Mihail Mintchev}
\medskip
\centerline {\it Istituto Nazionale di Fisica Nucleare, Sezione di Pisa}
\centerline {\it Dipartimento di Fisica dell'Universit\`a di Pisa,}
\centerline {\it Via Buonarroti 2, 56127 Pisa, Italy}
\bigskip
\medskip
\centerline {\large \rm Luigi Pilo}
\medskip
\centerline {\it Scuola Normale Superiore,}
\centerline {\it Piazza dei Cavalieri 7, 56126 Pisa, Italy} 
\medskip
\vskip 0.9 truecm
\centerline {\large \it Abstract} 
\medskip
\medskip 

An operator formalism for bosonization at finite temperature and density 
is developed. We treat the general case of anyon statistics. The exact 
$n$-point correlation functions, satisfying the Kubo-Martin-Schwinger condition, 
are explicitly constructed. The invariance under both vector and chiral 
transformations allows to introduce two chemical potentials. Investigating 
the exact momentum distribution, we discover anyon condensation in certain range 
of the statistical parameter. Another interesting feature is the occurrence of a 
non-vanishing persistent current. As an application of the general formalism, 
we solve the massless Thirring model at finite temperature, deriving 
the charge density and the persistent current. 

%%%%%%%%%%%%%

\bigskip
\noindent PACS numbers: 11.10Wx, 05.30Pr 
\medskip

\noindent Keywords: finite temperature field theory, anyons, condensation

\bigskip 

\centerline {May 1999}
\vfill \eject

\section{\bf Introduction}

Bosonization has become over the years an important tool for studying
various two dimensional field theory phenomena. It has been
successfully applied to the Tomonaga-Luttinger liquid model \cite{TL},
to quantum impurity problems \cite{HS} as well as in the theory of quantum
Hall edge states \cite{La}, \cite{We}. The possibility of describing in this
framework
also anyonic type excitations, attracts recently much attention.
Apart from a few considerations \cite{LMD}, \cite{LM}, generalized statistics
were not very popular in the past. The discovery of the quantum Hall effect
however, drastically changed the status of the subject, promoting it
into a fundamental concept of modern condensed matter theory.

In the physical applications of bosonization, one has to deal essentially
with systems
at finite temperature and density. This problem has been usually studied by
means
of functional integration in the real or imaginary time formalism of thermal
field
theory (see for instance \cite{Da}). In spite of the progress reached along
this way however, some relevant 
structural questions (e.g. the treatment of generalized statistics)
are still open. There exist also some controversies, presumably related
with genuine subtleties which appear in handling time-ordered paths
in the complex time-plane. For all these reasons, we find important
at this stage to test at work an alternative approach to bosonization
at finite temperature and density - the real time operator formalism.
This is precisely the goal of the present paper. We will work directly in
infinite volume, thus avoiding some delicate problems related with the
thermodynamic limit, especially in the presence of generalized statistics. 
The strategy will be to
construct explicitly the thermal correlation functions of anyons,
vector and chiral currents and to extract information about some physical
quantities like the momentum distribution, the density and the persistent
current.
Our framework sheds new light on the role of the vector and chiral
symmetries, which allow to introduce two independent chemical potentials. 
We expect the validity of our results to be quite general. Indeed, 
there are several indications (\cite{Vo}, \cite{WY}, \cite{Ha}) 
that the correlation functions we derive, capture 
the main universal features of one-dimensional gap-less quantum liquids, 
which support anyonic statistics and generalize the Tomonaga-Luttinger 
one \cite{To}, \cite{Lu}.

The paper is organized as follows. In Section 2 we collect some general
algebraic properties of the free boson field and its dual, describing also
the corresponding thermal representation. We also construct a set of
anyon fields and establish their general properties. Section 3 is devoted
to the derivation of the anyonic correlation functions at finite tempera\-ture
and density. The symmetry content of these correlators, which satisfy the
Kubo-Martin-Schwinger condition, is investigated. The momentum distribution
of the left- and right-moving anyonic modes is determined and a 
remarkable condensation-like phenomenon is detected. 
In Section 4 we consider the bosonization of fermion fields. A general family of
solutions of the massless Thirring model, comprising also anyonic solutions, 
is considered in Section 5. We determine both the density and the persistent
current of the model at finite temperature. Our derivation is free from any
ambiguity and resolves the existing discrepancies in the literature about
the coupling constant dependence of the density. The last section
summarizes the main results and contains our conclusions.

\section{\bf From bosons to anyons}

The fundamental building blocks for bosonization in 1+1 dimensions
are the free massless scalar field $\sci $ and its dual $\dsc $.
These fields are required to satisfy the equations of motion
\beeq
g^{\mu \nu}\der_\mu \der_\nu \, \sci (x) =
g^{\mu \nu}\der_\mu \der_\nu \, \dsc (x) = 0 \, ,
\qquad g = {\rm diag}\, \, (1,\, -1) \, ,
\label{eqm}
\end{equation}
the relations
\beeq
\der_\mu \dsc (x) = \varepsilon_{\mu \nu }\, \der^\nu \sci (x) \, ,
\qquad \varepsilon_{01} = 1 \, ,
\label{dual}
\end{equation}
and the equal-time canonical commutators
\beeq
[\sci (x)\, ,\, \sci (y)]\vert_{x^0=y^0} =
[\dsc (x)\, ,\, \dsc (y)]\vert_{x^0=y^0} = 0
\, ,
\label{ecc1}
\end{equation}
\beeq
[\der_0\sci (x)\, ,\, \sci (y)]\vert_{x^0=y^0} =
[\der_0\dsc (x)\, ,\, \dsc (y)]\vert_{x^0=y^0} =
-i\delta (x^1-y^1) \, .
\label{ecc2}
\end{equation}
Before constructing the finite temperature representation of $\sci $
and $\dsc $, it is useful to recall some of their basic features 
which are representation independent (see also \cite{St}, \cite{BLOT} and
\cite{AMS}).

\subsection{\bf General algebraic aspects}

One easily verifies that
\bea
&&\sci (x) = \frac{1}{ \sqrt 2} \int \mis
\left [a^\ast (k) \ep + a(k)\emi\right ] \quad ; \label{fil1} \\
&& \dsc (x) = \frac{1}{ \sqrt 2} \int \mis \varepsilon (k)
\left [a^\ast (k) \ep + a(k) \emi\right ] \quad , \label{fil2}
\ena
obey Eqs. (\ref{eqm},\ref{dual}). Here $\{a^\ast (k), a(k) \}$ are the
generators
of an associative algebra $\A_-$ with involution ${}^*$ and
identity element $\1$. The fields (\ref{fil1}-\ref{fil2})
satisfy the commutation relations (\ref{ecc1}-\ref{ecc2}) provided that
\bea
&&[a(k)\, ,\, a(p)] = [a^\ast (k)\, ,\, a^\ast (p)] = 0
\quad , \\
&& [a(k)\, ,\, a^\ast (p)] = \Delta (k) \, 2\pi \delta (k-p)\, \1
\label{aas}
\quad ,
\ena
$\Delta (k)$ being a solution of
\beeq
|k| \Delta (k) = 1 \quad .
\label{idef}
\end{equation}
There exists \cite{GS} a one-parameter family of distributions
$\{|k|^{-1}_\lambda \, :\, \lambda > 0 \}$, which obey
Eq. (\ref{idef}). A convenient representation is
\beeq
\int dk |k|^{-1}_\lambda f(k) \equiv
\lim_{\eta \downarrow 0}\, \frac{d}{ d \eta }\, \,  \eta
\left (\frac{\e^{\gamma_{{}_E}}}{ \lambda }\right )^\eta
\int dk |k|^{\eta -1} f(k) \quad ,
\end{equation}
where $f$ is a test function of rapid decrease and $\gamma_{{}_E}$ is
Euler's constant. The infrared origin of
the mass parameter $\lambda $ is clear and well understood \cite{Wa}.

A direct consequence of Eqs. (\ref{fil1}-\ref{idef}) are the commutation
relations
\bea
&&[\sci (x_1)\, ,\, \sci (x_2)] =
[\dsc (x_1)\, ,\, \dsc (x_2)] =
iD(x_1-x_2) \, \1
\quad ,  \label{cr1} \\
&& [\sci (x_1)\, ,\, \dsc (x_2)] =
i{\widetilde D}(x_1-x_2) \, \1 \label{cr2}
\quad ,
\ena
where
\bea
&&D(x) = -\frac{1}{ 2} \varepsilon (x^0)\, \theta (x^2) \, ,  \label{df1} \\
&& {\widetilde D}(x) =
\frac{1}{ 2} \varepsilon (x^1)\, \theta (-x^2) \, , \label{df2}
\ena
are $\lambda $-independent.
Eqs. (\ref{cr1},\ref{df1}) imply that $\sci $ and $\dsc $ are {\it local}
fields.
According to eqs. (\ref{cr2},\ref{df2}), $\sci $ and $\dsc $ are however
not {\it relatively local}. It is worth stressing that this
fact, known to be crucial for bosonization,
generalized (anyon) statistics, duality and chiral super-selected sectors,
is universal and does not depend on the representation of $\A_-$.
This observation is essential for finite temperature field theory,
where instead of the standard Fock representation $\F$ of $\A_-$, one
adopts a thermal representation $\T_-$ (see sect. 2.2), induced by a
Kubo-Martin-Schwinger (KMS) state \cite{BR} over $\A_-$.

In what follows we will frequently use the basis of chiral right
and left fields. In this basis locality is less transparent, but one gains
nice factorization properties. Introducing the light-cone coordinates
\beeq
x^\pm = x^0 \pm x^1 \quad ,
\end{equation}
the chiral fields are defined by
\bea
&&\scr (x^-) \equiv
{\sqrt 2} \int_0^\infty \mis \left [a^\ast (k) \e^{ikx^-} +
a(k)\e^{-ikx^-} \right ] = \sci (x) + \dsc (x) \,  , \label{lf} \\
&& \scl (x^+) \equiv
{\sqrt 2} \int^0_{-\infty }\mis \left [a^\ast (k) \e^{-ikx^+} +
a(k) \e^{ikx^+}\right ] =
\sci (x) - \dsc (x) \, . \label{rf}
\ena
Combining eqs. (\ref{cr1}-\ref{df2},\ref{lf},\ref{rf}), one finds
\beeq
[\sczu (\zeta_1 )\, ,\, \sczd (\zeta_2 )] =
-i\, \varepsilon (\zeta_{12})\, \dlt \, \1
\, , \qquad \zeta_{12} \equiv \zeta_1 - \zeta_2 \quad , \label{cz}
\end{equation}
where $Z = L, R$ and
\beeq
\dlt = \left\{ \begin{array}{cc} 1 & \mbox{if } Z_1=Z_2 \\
0 & \mbox{if } Z_1 \neq Z_2
\end{array} \right.
\end{equation}
Let us denote by $\az$ the associative algebra generated
by $\scz $. Given any smooth real valued function $\fz $, 
it is easily checked that the shift transformation
\beeq
\alpha_{\fz }\, :\, \scz (\zeta ) \longmapsto
\scz (\zeta ) + \fz (\zeta ) \, \1 \,  ,
\label{saut}
\end{equation}
leaves invariant both eqs. (\ref{eqm},\ref{dual}) and the
commutation relations (\ref{cz}). Therefore, $\alpha_{f_Z}$ is an automorphism
of $\az$. It can be given the form
\beeq
\alpha_{\fz} [\scz (\zeta )] = \e^{i \cqz }\, \scz (\zeta )\, \e^{-i \cqz }
\quad ,
\end{equation}
where the charges $\cqz $ satisfy
\bea
&&[\cqzu  \, , \, \sczu (\zeta )] = -i f_{{}_{Z_1}} (\zeta ) \, \dlt \1\, ,
\label{char1} \\
&&[\cqzu \, ,\, \cqzd ] = 0 \, . \label{char2}
\ena
The automorphism (\ref{saut}) turns out to be essential for bosonization with
non-vanishing chemical potential.

We now turn to generalized statistics, which can be discussed 
at purely algebraic level. In fact, let us concentrate on the set of
fields
parametrized by $\xi= (\sigma,\, \tau ) \in \R^2$ and defined by
\beeq
A(x;\xi) \equiv
z(\lambda ;\xi)\, \exp \left [\frac{i\pi} {2} \left (\tau \qr - \sigma \ql
\right )\right ]
:\exp \left \{i\sqrt \pi \left [\sigma \scr (x^-) +
\tau \scl (x^+) \right ]\right \}: \, ,
\label{adef}
\end{equation}
where
\beeq
\qz \equiv \frac{1}{ \sqrt \pi}{\Q}_{{}_Z}(1)\, ,
\label{lq}
\end{equation}
and the normal ordering $\, : \quad  :\, $ is taken with respect to
the creators $a^*(k)$ and annihilators $a(k)$. The normalization
$z(\lambda ;\xi)$ will be fixed in what follows. Let us observe
that due to eqs. (\ref{char1},\ref{char2}), the relative position of the two
exponential factors in (\ref{adef}) is irrelevant.
Notice also that under $*$-conjugation,
\beeq
A^*(x;\xi) = \frac{{\overline z}(\lambda ;\xi)}{ z(\lambda ;-\xi)}\, A(x;-\xi)
\quad .
\label{acon}
\end{equation}
Although one may consider expressions more general then (\ref{adef}),
the field $A(x;\zeta )$ will be enough for our purposes.

Using eqs. (\ref{cz},\ref{char1},\ref{char2}), we derive the exchange
relation
\beeq
A(x_1;\xi_1 ) A(x_2;\xi_2 ) =
{\cal R}(x_{12};\xi_1, \xi_2 )\,
A(x_2;\xi_2 ) A(x_1;\xi_1 )
\, .
\end{equation}
The exchange factor $\cal R$ reads
\beeq
{\cal R}(x_{12};\xi_1, \xi_2 ) =
\exp \bigl \{ - i\pi \bigl [2(\sigma_1 \sigma_2 + \tau_1 \tau_2)D(x_{12}) +
2(\sigma_1 \sigma_2 - \tau_1 \tau_2){\widetilde D}(x_{12}) +
(\sigma_1 \tau_2 - \sigma_2 \tau_1) \bigr ] \bigr \} .
\label{rdef}
\end{equation}
The statistics of the field (\ref{adef}) is governed by the behavior
of (\ref{rdef}) for $\xi_1 = \xi_2 = \xi$ and space-like separated
points $x_{12}^2 < 0$. By means of eqs. (\ref{df1},\ref{df2}), one gets
in this domain
\beeq
A(x_1;\xi) A(x_2;\xi) =
\exp [-i\pi (\sigma^2 - \tau^2) \varepsilon (x^1_1-x^1_2)]
\, A(x_2;\xi) A(x_1;\xi) \quad .
\label{exc}
\end{equation}
Therefore, $A(x;\xi)$ is an anyon field whose statistics parameter is
\beeq
\vartheta (\xi ) = \sigma^2 - \tau^2 \quad .
\label{stat}
\end{equation}
Bose or Fermi statistics are recovered when $\vartheta $ is an even or odd
integer respectively. The remaining values of $\vartheta $ lead to
Abelian braid statistics. The basic fact behind this implementation of 
generalized statistics, is clearly the relative non-locality of
$\sci $ and $\dsc $.

In our discussion below we frequently use the chiral currents
\beeq
\jz (\zeta ) \equiv \frac{1}{ \sqrt \pi}\, \der \scz (\zeta )  \quad .
\label{bcurr}
\end{equation}
A straightforward computation, based on eq. (\ref{cz}), gives
\beeq
[j_{{}_Z}(\zeta ) \, , \, A(x;\xi) ] = 2\xi^Z \delta (\zeta - x^Z) A(x;\xi)
\, ,
\label{abc}
\end{equation}
where
\beeq
\xi^Z = \left\{ \begin{array}{cc} \sigma  & \mbox{if } Z=R \\
\tau  & \mbox{if } Z=L \end{array} \right. \qquad
x^Z =  \left\{ \begin{array}{cc} x^- & \mbox{if } Z=R  \\
x^+  & \mbox{if } Z=L \end{array} \right.
\label{not}
\end{equation}
Moreover, differentiating (\ref{adef}) with respect to $x^Z$, one finds
\beeq
\frac{\der}{ \der x^Z} A(x;\xi) =
i \pi \xi^Z : j_{{}_Z}(x^Z) \, A(x; \xi ) : \, .
\label{da}
\end{equation}

The energy-momentum tensor of the fields $\sci $ and $\dsc $ has two
independent
components, which read in the chiral basis
\beeq
\Theta_{{}_Z}(\zeta ) = \frac{1}{ 4}:\der \scz (\zeta ) \, \der \scz (\zeta ) :
= \frac{\pi}{ 4} :\jz (\zeta )\, \jz (\zeta): \, .
\label{bem}
\end{equation}
Indeed, one can directly verify that
\beeq
[\Theta_{{}_Z}(\zeta_1) \, , \, \scz(\zeta_2) ]= - i \der
\scz (\zeta_2)\, \delta(\zeta_{12}) \, .
\label{tf}
\end{equation}

{}Finally, we would like to point out that the short distance expansion of the
product
$A^*(t,x_1;\xi) A(t,x_2;\xi)$ involves the chiral currents. In fact, normal
ordering this product, one finds
\bea
&&A^*(x_1;\xi)\, A(x_2;\xi) =
|z(\lambda ;\xi )|^2\, \exp \left [\pi \sigma^2 u(\lambda x_{12}^-) +
\pi \tau^2 u(\lambda x_{12}^+)\right ] \cdot \nb \\
&& : \exp \left \{ i\sqrt \pi \left [\sigma \scr (x_2^-) - \sigma \scr (x_1^-)
+ \tau \scl (x_2^+) - \tau \scl (x_1^+) \right ] \right \} : \; ,
\label{aa}
\ena
where
\beeq
u(\zeta ) = - \frac{1}{ \pi } \ln (i\zeta + \epsilon )
= -\frac{1}{ \pi } \ln |\zeta | - \frac{i}{ 2}\varepsilon (\zeta )
\, ,
\label{uf}
\end{equation}
the presence of the parameter $\epsilon >0$ implies as usual
the weak limit $\epsilon \to 0$. From (\ref{aa}) one derives the expansion
\bea
&&C(x,\eta;\xi) \equiv
\frac{1}{ 2}\left [A^*(x + \eta ;\xi)\, A(x;\xi) - A(x;\xi)\, A^*(x-\eta
;\xi)\right ] = \nb \\
&&
i\pi |z(\lambda ;\xi )|^2\, \e^{\pi [\sigma^2 u(\lambda \eta^-) +
\tau^2 u(\lambda \eta^+)]} \,
\left [\sigma \eta^- j_{{}_R}(x^-) +
\tau \eta^+ j_{{}_L}(x^+)
+ O(\eta^+) + O(\eta^-)\right ] \, .
\label{xxx}
\ena
Eqs. (\ref{abc},\ref{da},\ref{tf}) are useful for
analyzing the symmetry content and the dynamics of $A(x;\xi)$,
whereas (\ref{xxx}) is convenient in the study of the short
distance behavior.

Summarizing, we have introduced the basic algebraic tools for
bosonization - the oscillator algebra $\A_-$ and the related field
algebras $\A_{{}_R}$ and $\A_{{}_L}$. We have established also 
some identities needed in the construction of fermions and, more generally,
anyons. Since $\A_-$ and $\A_{{}_Z}$ are infinite algebras, some of our
manipulations above were formal. In what follows however, we shall deal
with operator representations of $\A_-$ and $\A_{{}_Z}$, in which the
just mentioned identities hold at least in mean value on a suitable
domain of the state space.

\subsection {\bf  Thermal representations}

We start this section by constructing a thermal representation $\T_-$ of
$\A_-$,
whose cyclic vector is the Gibbs (grand canonical) equilibrium state
associated with
\beeq
K_{{}_B} \equiv H_{{}_B} - \mu_{{}_B} N_{{}_B} \, , \qquad \mu_{{}_B} < 0
\, .
\label{chem}
\end{equation}
Here
\beeq
H_{{}_B} = \int \mis k^2\, a^\ast (k) a(k) \, , \qquad
N_{{}_B} = \int \mis |k| \, a^\ast (k) a(k) \, ,
\end{equation}
are the Hamiltonian and the number operator respectively, and
$\mu_{{}_B}$ is the chemical potential. The non-vanishing two-point
correlation functions in this state are given by
\bea
&&\langle a^*(p)a(q)\rangle_{\mu_{{}_B}}^\beta =
\frac{\e^{-\beta (|p|-\mu_B)}}{ 1-\e^{-\beta (|p|-\mu_B)}} \,
|p|^{-1}_{\lambda }\, 2\pi \delta (p-q) \, , \; \label{be1} \\
&&
\langle a(q)a^*(p)\rangle_{\mu_B}^\beta =
\frac{1}{ 1-\e^{-\beta (|p|-\mu_B)}}\,
|p|^{-1}_{\lambda }\, 2\pi \delta (p-q) \, ,
\label{be2}
\ena
$\beta $ being the inverse temperature. Notice that the Bose-Einstein 
distribution appears as a factor in the right hand side of (\ref{be1},\ref{be2}). 
The parameter $\mu_{{}_B} < 0$ allows to work with well defined bosonic 
correlators. Dealing with the physical anyon fields, introduced in 
the next section, we will take the limit $\mu_{{}_B} \to 0$. Let us also 
recall for comparison that
\beeq
\langle a^*(p)a(q)\rangle_{{}_F} = 0 \, , \qquad
\langle a(q)a^*(p)\rangle_{{}_F} =
|p|^{-1}_{\lambda }\, 2\pi \delta (p-q) \, .
\label{fock}
\end{equation}
in the Fock representation $\F $ of $\A_-$.

Referring to \cite{BR} for the rigorous argument, it is useful to
sketch here a heuristic derivation of the expectation values 
(\ref{be1},\ref{be2}). Since they are constrained by the commutator (\ref{aas}),
it is enough to concentrate for instance on
eq. (\ref{be1}). Setting
\beeq
\langle a^*(p)a(q)\rangle_{\mu_{{}_B}}^\beta \equiv
\frac{{\rm Tr}\left [\e^{-\beta K_{{}_B}}a^*(p)a(q)\right ]}{
{\rm Tr}\, \e^{-\beta K_{{}_B}}} \, ,
\end{equation}
one finds
\bea
&&\langle a^*(p)a(q)\rangle_{\mu_B}^\beta =
\frac{{\rm Tr}\left [\e^{-\beta K_{{}_B}}a^*(p)a(q)\right ]}{ 
{\rm Tr}\, \e^{-\beta K_{{}_B}}} = \nb \\
&&
\e^{-\beta (|p|-\mu_{{}_B})}
\frac{{\rm Tr}\left [a^*(p)\e^{-\beta K_{{}_B}}a(q)\right ]}{
{\rm Tr}\, \e^{-\beta K_{{}_B}}} =
\e^{-\beta (|p|-\mu_{{}_B})}
\frac{{\rm Tr}\left [\e^{-\beta K_{{}_B}}a(q)a^*(p)\right ]}{
{\rm Tr}\, \e^{-\beta K_{{}_B}}} = \nb \\
&&
\e^{-\beta (|p|-\mu_{{}_B})}\, |p|_\lambda^{-1}\, 2\pi \delta (p-q) +
\e^{-\beta (|p|-\mu_{{}_B})} \langle a^*(p)a(q)\rangle_{\mu_{{}_B}}^\beta
\, ,
\label{qua}
\ena
where eq.(\ref{aas}) and the cyclicity of the trace have been used. 
Eq. (\ref{qua}) can be solved for $\langle a^*(p)a(q)\rangle_{\mu_{{}_B}}^\beta
$ and gives
precisely (\ref{be1}).

Using the commutation relation (\ref{aas}), a generic correlation function
can be reduced to
\beeq
\langle \prod_{i=1}^m a^*(p_i) \prod_{j=1}^n a(q_j)
\rangle_{\mu_{{}_B}}^\beta \, ,
\end{equation}
which can be evaluated by iteration from (see e.g. \cite{BR})
\beeq
\langle \prod_{i=1}^m a^*(p_i) \prod_{j=1}^n a(q_j)
\rangle_{\mu_{{}_B}}^\beta =
\delta_{mn}\, \sum_{k=1}^m \langle a^*(p_1)a(q_k)\rangle_{\mu_{{}_B}}^\beta
\, \langle \prod_{i=2}^m a^*(p_i) \prod_{\stackrel{j=1}{j\not=k} }^n a(q_j)
\rangle_{\mu_{{}_B}}^\beta \, \, ,
\label{2a}
\end{equation}
and the normalization condition
\beeq
\langle \1 \rangle_{\mu_{{}_B}}^\beta = 1 \, . 
\label{nc}
\end{equation}
As it should be expected, the correlators derived in this way
satisfy the KMS condition corresponding to the automorphism of $\A_-$ 
\beeq
\alpha_s \, a^*(k) = a^*(k) \e^{is(|k|- \mu_{{}_B} )} \, ,\qquad
\alpha_s \, a(k) = a(k) \e^{-is(|k|- \mu_{{}_B} )} \, ,
\end{equation}
generated by $K_{{}_B} $. One easily checks for example that
\beeq
\langle \left [\alpha_s a(q)\right ]a^*(p)\rangle_{\mu_{{}_B}}^\beta =
\langle a^*(p)\left [\alpha_{s+i\beta} a(q)\right
]\rangle_{\mu_{{}_B}}^\beta \, .
\end{equation}

Summarizing, eqs. (\ref{be1},\ref{be2},\ref{2a},\ref{nc}) completely determine the
thermal representation $\T_-$ of $\A_-$.
Since the distribution $|k|_\lambda^{-1}$ is not positive definite,
$\T_-$ has indefinite metric. This property is peculiar of the
two-dimensional world and it is also present in the Fock representation 
\cite{M} defined by eq. (\ref{fock}). We will address some of its aspects 
later on, when discussing the physical representation of the anyon field 
(\ref{adef}).

It is straightforward at this point to compute the correlation functions
of $\scr $ and $\scl $. The mixed correlators vanish and we are left with
$\langle \scz (\zeta _1)\cdots \scz (\zeta _m) \rangle_{\mu_{{}_B}}^\beta $,
which define a thermal representation $\T_{{}_Z}$ of the algebra $\A_{{}_Z}$.
One finds,
\beeq
\langle \scz (\zeta _1)\cdots \scz (\zeta _{2n+1})
\rangle_{\mu_{{}_B}}^\beta = 0 \, , \label{sel}
\end{equation}
\beeq
\langle \scz (\zeta _1)\cdots \scz (\zeta _{2n}) \rangle_{\mu_{{}_B}}^\beta =
\sum \langle \scz (\zeta _{k_1}) \scz (\zeta _{k_2})
\rangle_{\mu_{{}_B}}^\beta \cdots
\langle \scz (\zeta _{k_{2n-1}}) \scz (\zeta _{k_{2n}})
\rangle_{\mu_{{}_B}}^\beta \, ,
\end{equation}
where the summation runs over all permutations of the integers $\{1,...,2n\}$,
such that\break $k_1<k_3<...<k_{2n-1}$ and $k_{2j-1}<k_{2j}$ for
$j=1,...,n$.
The two-point function, which is the fundamental building block, reads
\bea
&&\langle \scz (\zeta _1) \scz (\zeta _2) \rangle_{\mu_{{}_B}}^\beta
\equiv w (\zeta _{12};\beta , \mu_{{}_B} ) = \nb \\
&&
2\int_0^\infty \mis \, |k|^{-1}_{\lambda } \left [
\frac{\e^{-\beta (|k|-\mu_{{}_B})}}{ 1-\e^{-\beta (|k|-\mu_{{}_B})}}\,
\e^{ik(\zeta _{12}-i\epsilon )}
+ \frac{1}{ 1-\e^{-\beta (|k|-\mu_{{}_B})}}\, \e^{-ik(\zeta _{12}-i\epsilon
)} \right ]
\, .
\label{irep}
\ena
Since $\lambda >0$ and $|\mu_{{}_B}|>0$,
eq. (\ref{irep}) provides a well defined integral representation of
$w (\zeta ; \beta ,\mu_{{}_B} )$. The full $\lambda $-dependence and
the singularity in $\mu_{{}_B} = 0$ are captured by
\beeq
w (\zeta ;\beta , \mu_{{}_B} ) =
\frac{1}{ \pi}\left \{ \frac{2}{ \beta |\mu_{{}_B} |}\ln \frac{|\mu_{{}_B}
|\e^{\gamma_{{}_E}}}{ \lambda }
- \ln \left [2i\, \sinh \left (\frac{\pi}{  \beta}\, \zeta - i\epsilon \right )
\right ] \right \} + O(\mu_{{}_B} ) \, ,
\label{cdep}
\end{equation}
where $O(\mu_{{}_B} )$ stands for $\lambda$-independent terms vanishing in the
limit $\mu_{{}_B} \to 0$.

As already mentioned, the parameter $\mu_{{}_B}$ has been introduced 
exclusively for technical reasons. It has nothing to do with the chemical 
potentials for fermions and anyons. In order to recover the latter, 
we need the automorphism 
\beeq
\alpha_{\mu_{{}_Z}}\, :\, \scz (\zeta ) \longmapsto \scz (\zeta ) -
\frac{\mu_{{}_Z}}{  \sqrt \pi}\, \zeta \qquad  \mu_{{}_Z} \in \R \, ,
\label{baut}
\end{equation}
of $\A_{{}_Z}$, obtained by setting
\beeq
\fz(\zeta ) = - \frac{\mu_{{}_Z}}{  \sqrt \pi}\, \zeta \, ,
\end{equation}
in eq. (\ref{saut}). By inspection, $\alpha_{\mu_{{}_Z}} $ can not be
implemented
unitarily in $\T_{{}_Z}$. In fact,
\beeq
\langle \alpha_{\mu_{{}_Z}} \left [\scz (\zeta _1) \scz (\zeta _2)\right ]
\rangle_{\mu_{{}_B}}^\beta
= \langle \scz (\zeta _1) \scz (\zeta _2) \rangle_{\mu_{{}_B}}^\beta +
\frac{\mu^2_{{}_Z}}{ \pi }\, \zeta _1 \zeta _2  \, .
\end{equation}
Therefore, by applying $\alpha_{\mu_{{}_Z}}$ one generates from $\T_{{}_Z}$
a new representation $\T_{{}_Z}(\mu_{{}_Z})$ of $\A_{{}_Z}$, associated with
the correlation functions
$\langle \alpha_{\mu_{{}_Z}} \left [\scz (\zeta _1)\cdots
\scz (\zeta _n)\right ] \rangle_{\mu_{{}_B}}^\beta $, which can be obtained
directly from eqs. (\ref{sel}-\ref{irep}) by the shift (\ref{baut}).

Concerning the correlators
$\langle \alpha_{\mu_{{}_Z}} \left [\jz (\zeta _1)\cdots
\jz (\zeta _n)\right ] \rangle_{\mu_{{}_B}}^\beta $ of the chiral
currents (\ref{bcurr}) in $\T_{{}_Z}(\mu_{{}_Z})$, it follows from eqs.
(\ref{sel}-\ref{cdep})
that they are both $\lambda $-independent and regular in the limit $\mu_{{}_B}
\to 0$.
Defining
\beeq
\langle \jz (\zeta _1)\cdots \jz (\zeta _m) \rangle_{\mu_{{}_Z}}^\beta \equiv
\lim_{\mu_{{}_B} \uparrow 0}\,
\langle \alpha_{\mu_{{}_Z}}
\left [\jz (\zeta _1)\cdots \jz (\zeta _m)\right ] \rangle_{\mu_{{}_B}}^\beta
\quad ,
\label{ccdef}
\end{equation}
one has for instance
\bea
&&\langle \jz (\zeta ) \rangle_{\mu_{{}_Z}}^\beta = - \frac{\mu_{{}_Z}}{  \pi}
\quad , \label{exj} \\
&&
\langle \jz (\zeta _1) \jz (\zeta _2)\rangle_{\mu_{{}_Z}}^\beta
= \frac{\mu^2_{{}_Z}}{  \pi^2} -
\frac{1}{\ \beta^2}\, \sinh^{-2} \left (\frac{\pi}{ \beta}\, \zeta _{12} -
i\epsilon \right )  \, .
\label{bc2p}
\ena
At this point, one can derive the correlation functions of
$\Theta_{{}_Z}(\zeta )$ from those
of the currents, using the following weak limit representation
\beeq
\Theta_{{}_Z} (\zeta)= \frac{1}{ 4}
\lim_{\zeta'\to\zeta}:\der\varphi_Z(\zeta')\der\varphi_Z(\zeta):\,  =
\frac{1}{ 4}\lim_{\zeta'\to\zeta}\left[\pi \jz (\zeta') \jz (\zeta)
+ \frac{1}{ \pi (\zeta'-\zeta-i\epsilon)^2}\right] \, ,
\label{lim}
\end{equation}
valid in $\T_{{}_Z}(\mu_{{}_Z})$. Combining (\ref{bc2p}) and (\ref{lim}), one
finds
\beeq
\langle \Theta_{{}_Z}(\zeta ) \rangle_{\mu_{{}_Z}}^\beta  =
\frac{\mu_{{}_Z}^2 }{ 4\pi} +
\frac{\pi}{ 12\beta^2}\, .
\label{eme}
\end{equation}
In conclusion, starting with the thermal representation $\T_-$ of $\A_-$,
we have constructed the representation $\T_{{}_Z}(\mu_{{}_Z})$ of
$\A_{{}_Z}$.
We will show in what follows that $\T_{{}_R}(\mu_{{}_R})$ and
$\T_{{}_L}(\mu_{{}_L})$
provide the basis for bosonization at finite temperature.

\section{\bf Anyons at finite temperature and density}
In this section we adopt the tensor product
$\T_{{}_L}(\mu_{{}_L}) \otimes \T_{{}_R}(\mu_{{}_R})$ for the construction
of anyonic fields.
\subsection{\bf The physical representation - basic properties and
correlation functions}
By analogy with the zero temperature case, we concentrate on the
$q_{{}_Z}$-symmetric
anyon correlation functions. Because of eq. (\ref{abc}), they are
characterized by two selection rules, namely 
\beeq
\sum_{i=1}^n \sigma_i = 0\, , \qquad  \sum_{i=1}^n \tau_i = 0 \, .
\label{sel1}
\end{equation}
Using (\ref{sel1}), after some algebra one finds
\bea
&&\qquad \qquad \qquad \langle \alpha_{\mu_{{}_Z}} \left [A(x_1;\xi_1)\cdots
A(x_n;\xi_n)\right ] \rangle_{\mu_{{}_B}}^\beta = \nb \\
&&
\left [\prod_{i=1}^n z(\lambda ;\xi_i )
\exp (-i\mu_{{}_R} \sigma_i x_i^- - i\mu_{{}_L}\tau_ix_i^+) \right ]
\, \exp \left [- \pi \, v(\beta ,\mu_{{}_B} ) \sum_{i=1}^n (\sigma_i^2 +
\tau_i^2)
\right ]   \cdot \nb \\
&&
  \exp \left \{-\pi\sum_{\stackrel{i,j = 1}{ i<j}}^n \left [
\sigma_i \sigma_j w (x_{ij}^-;\beta ,\mu_{{}_B} ) +
\tau_i \tau_j w (x_{ij}^+;\beta ,\mu_{{}_B} ) -
\frac{i}{ 2} \left (\tau_i \sigma_j - \tau_j \sigma_i \right )\right ] \right \}
\, ,
\label{lc}
\ena
where
\beeq
v(\beta ,\mu_{{}_B} ) \equiv 2 \int_0^\infty \mis
|k|_\lambda^{-1}\, \frac{\e^{-\beta (|k|-\mu_{{}_B})}}{ 1-\e^{-\beta
(|k|-\mu_{{}_B})}} =
\frac{1}{ \pi} \left ( \frac{1}{ \beta |\mu_{{}_B} |}\ln {|\mu_{{}_B}
| \frac{\e^{\gamma_{{}_E}}}{ \lambda }}
+ \frac{1}{ 2} \ln \frac{\beta \lambda}{  2\pi }
\right )  + O(\mu_{{}_B} ) \, .
\end{equation}
As in eq. (\ref{cdep}), $O(\mu_{{}_B} )$ denotes $\lambda$-independent terms
which vanish for $\mu_{{}_B} \to 0$. Conditions (\ref{sel1})
immediately imply
translation invariance of (\ref{lc}). A closer inspection reveals also that
the right hand side of eq. (\ref{lc}) is:
\begin{itemize}
\item {(i)} $\lambda $-independent;
\item {(ii)} well defined in the limit $\mu_{{}_B} \to 0\, $,
\end{itemize}
provided that
\beeq
z(\lambda ;\xi) =
\left (\frac{\lambda}{ 2\pi}\right )^{\frac{1}{ 2}(\sigma^2 + \tau^2)} \quad .
\label{ren}
\end{equation}
These features are shared also by the correlation functions
\beeq
\langle \alpha_{\mu_{{}_Z}} \left [j_{{}_{Z_1}}(\zeta_1)\cdots
j_{{}_{Z_m}}(\zeta_m)
A(x_1;\xi_1)\cdots A(x_n;\xi_n)\right ] \rangle_{\mu_{{}_B}}^\beta \, ,
\end{equation}
involving the chiral current (\ref{bcurr}). Therefore, from now on, we impose 
both (\ref{sel1}) and (\ref{ren}) and define 
\bea
&&\langle j_{{}_{Z_1}}(\zeta_1)\cdots j_{{}_{Z_m}}(\zeta_m)A(x_1;\xi_1)\cdots
A(x_n;\xi_n )
\rangle_{\mu_{{}_L},\, \mu_{{}_R}}^\beta \equiv \nb \\
&&
\lim_{\mu_{{}_B} \uparrow 0}\, \langle \alpha_{\mu_{{}_Z}}
\left [j_{{}_{Z_1}}(\zeta_1)\cdots j_{{}_{Z_m}}(\zeta_m)
A(x_1;\xi_1)\cdots A(x_n;\xi_n)\right ]\rangle_{\mu_{{}_B}}^\beta \, .
\label{cacd}
\ena
Eq. (\ref{lc}) implies
\bea
&&\langle A(x_1;\xi_1)\cdots A(x_n;\xi_n) \rangle_{\mu_L,\, \mu_R}^\beta =
\nb \\
&&
\exp \left [\frac{i \pi}{2} \sum_{\stackrel{i,j=1}{i<j}}^n (\tau_i \sigma_j - \tau_j
\sigma_i )
-i\mu_{{}_R}\sum_{i=1}^n \sigma_i x_i^- - i\mu_{{}_L} \sum_{i=1}^n
\tau_i x_i^+ \right ] \left (\frac{1}{ 2 \pi} \right )^{\frac{1}{2}\sum_{i=1}^n
(\sigma_i^2 + \tau_i^2)}  \nb \\
&&
\cdot \prod_{\stackrel{i,j=1}{ i<j}}^n
\left [ \frac{i \beta}{ \pi } \sinh
\left (\frac{\pi}{ \beta} x_{ij}^- -i\epsilon \right )\right ]^{\sigma_i
\sigma_j}
\left [ \frac{i \beta}{  \pi } \sinh
\left (\frac{\pi }{ \beta}x_{ij}^+ -i\epsilon \right )\right ]^{\tau_i \tau_j}
\, \label{ac}
\ena
Then, a generic correlator of the type (\ref{cacd}) can be obtained by
iteration via
\bea
&&\langle j_{{}_{Z_1}}(\zeta_1)j_{{}_{Z_2}}(\zeta_2)
\cdots j_{{}_{Z_m}}(\zeta_m)A(x_1;\xi_1)\cdots  A(x_n;\xi_n )
\rangle_{\mu_{{}_L},\, \mu_{{}_R}}^\beta = \nb \\
&&-\left \{ \frac{\mu_{{}_{Z_1}} }{ \pi} + \frac{i}{ \beta} \sum_{j=1}^n
\xi_j^{Z_j}
\coth \left [ \frac{\pi}{ \beta}(\zeta_1 - x_j^{Z_1}) - i\epsilon \right ]
\right \} \nb \\
&&
\cdot \, \langle j_{{}_{Z_2}}(\zeta_2)\cdots
j_{{}_{Z_m}}(\zeta_m)A(x_1;\xi_1)\cdots  A(x_n;\xi_n )
\rangle_{\mu_{{}_L},\, \mu_{{}_R}}^\beta
- \frac{1}{ \beta^2}\sum_{j=2}^m \dlt \sinh^{-2}\left (\frac{\pi}{\beta }
\zeta_{1j} - i\epsilon \right ) \nb \\
&&\cdot \langle j_{{}_{Z_2}}(\zeta_2)\cdots \widehat {j_{{}_{Z_j}}}(\zeta_j)
\cdots j_{{}_{Z_m}}(\zeta_m)A(x_1;\xi_1)\cdots  A(x_n;\xi_n )
\rangle_{\mu_{{}_L},\, \mu_{{}_R}}^\beta \, ,
\label{jac}
\ena
the hat in the right hand side of (\ref{jac}) denotes that the
corresponding current is omitted. One has for instance,
\bea
&& \langle j_{{}_R}(\zeta_1)j_{{}_R}(\zeta_2) A(x_1;\xi_1)\cdots  A(x_n;\xi_n )
\rangle_{\mu_{{}_L},\, \mu_{{}_R}}^\beta = \nb \\
&&\langle A(x_1;\xi_1)\cdots A(x_n;\xi_n) \rangle_{\mu_L,\, \mu_R}^\beta \cdot
\nb \\
&&
\Biggl \{ \frac{\mu^2_{{}_R}}{  \pi^2} -
\frac{1}{ \beta^2}\, \sinh^{-2} \left (\frac{\pi}{  \beta}\, \zeta_{12} -
i\epsilon \right )
+i \frac{\mu_{{}_R}}{  \pi \beta} \sum_{j=1}^n \sigma_j \sum_{k=1}^2
\coth \left [\frac{\pi}{  \beta}\, (\zeta_k-x_j^-) - i\epsilon \right ] \nb \\
&&
- \frac{1}{  \beta^2} \sum_{j,k=1}^n \sigma_j \sigma_k \,
\coth \left [\frac{\pi}{  \beta}\, (\zeta_1 - x_j^-) - i\epsilon \right ]
\coth \left [\frac{\pi}{  \beta}\, (\zeta_2 - x_k^-) - i\epsilon \right ]
\Biggr \} \, .
\label{ex}
\ena

Coming back to the points (i-ii) above, we observe that
the property (i) is essential for our construction.
Like at zero temperature, the $\lambda $-independence of (\ref{ac}) signals
positivity. In fact, repeating with some obvious modifications the argument of
reference \cite{pos}, one can prove that the correlation functions
(\ref{cacd}) are
positive definite, which is crucial for the physical interpretation.
Point (ii) in turn, allows to eliminate $\mu_{{}_B}$ from the
$\qz $-symmetric correlators, in agreement with the fact that the 
relevant chemical potentials at anyonic level are $\mu_{{}_Z}$. 
This issue is discussed few lines below.

We are now in a position to perform an important step -
the definition of the physical quantum fields $\{A^{\rm ph}(x;\xi ),\,
j^{\rm ph}_{{}_Z}(\zeta )\}$.
Let us introduce for this purpose the set\break $\Xi = \{\xi_1,
-\xi_1,...,\xi_n, -\xi_n \}$
and let us consider the family $\{ A(x;\xi )\, :\, \xi \in \Xi\}$. We define
\bea
&&\langle j_{{}_{Z_1}}^{\rm ph}(\zeta_1)\cdots j_{{}_{Z_m}}^{\rm ph}(\zeta_m)
A^{\rm ph}(x_1;\xi_1)\cdots  A^{\rm ph}(x_n;\xi_n )
\rangle_{\mu_{{}_L},\, \mu_{{}_R}}^{{\rm ph},\, \beta }\equiv \nb \\
&&\left \{ \begin{array}{cc}
\langle j_{{}_{Z_1}}(\zeta_1)\cdots j_{{}_{Z_m}}(\zeta_m)A(x_1;\xi_1)\cdots
A(x_n;\xi_n )
\rangle_{\mu_{{}_L},\, \mu_{{}_R}}^\beta
& \mbox{if (\ref{sel1}) holds;} \\
0 & \mbox{otherwise.} \end{array} \right.
\label{phd}
\ena
Being positive definite by construction, the functions (\ref{phd})
determine, via the reconstruction theorem \cite{BLOT}, the fields
$\{ A^{\rm ph}(x;\xi )\, , j^{\rm ph}_{{}_Z}(\zeta )\, :\, \xi \in \Xi\}$.
The explicit form of the physical correlators (\ref{phd}) is quite
remarkable and
deserves a comment. Without current insertions, the equal-time $n$-point
function is
a finite temperature and density generalization of the Jastrow-Laughlin
wave function
\cite{La}. We recall that the latter  describes the $n$-particle ground
state in several
one-dimensional models, showing a Tomonaga-Luttinger liquid structure. In that 
context,
the current insertions in (\ref{phd}) are associated with the charged
excitations of
the liquid. Let us mention also that (\ref{phd}) admits a well defined
continuation to 
imaginary time, which still represents finite temperature
and density version of the Jastrow-Laughlin wave function, though in
two spatial dimensions.

The expectation values (\ref{phd}) are invariant under the transformations:
\bea
&&A^{\rm ph}(x ;\xi ) \longmapsto A^{\rm ph}((x^0+t ,x^1);\xi )\, ,
\qquad j^{\rm ph}_{{}_Z}(\zeta ) \longmapsto j^{\rm ph}_{{}_Z}(\zeta + t)\,
, \qquad
t \in \R \, ; \\
&&A^{\rm ph}(x;\xi ) \longmapsto \e^{-is_{{}_R}\sigma }\, A^{\rm ph}(x;\xi )
\, ,
\qquad \quad \, \, j^{\rm ph}_{{}_Z}(\zeta ) \longmapsto j^{\rm ph}_{{}_Z}(\zeta )\, ,
\qquad \quad s_{{}_R}\in  \R \, ; \\
&&A^{\rm ph}(x;\xi ) \longmapsto \e^{-is_{{}_L}\tau }\, A^{\rm ph}(x;\xi ) \, ,
\qquad \quad \, \, j^{\rm ph}_{{}_Z}(\zeta ) \longmapsto j^{\rm ph}_{{}_Z}(\zeta )\, ,
\qquad \quad s_{{}_L} \in \R \, .
\ena
We denote the corresponding conserved charges by $H$ and $\Qz$ and 
we consider the automorphism $\alpha_s$ generated by
\beeq
K \equiv H -\mu_{{}_L}\Ql - \mu_{{}_R}\Qr \quad .
\end{equation}
Using that
\beeq
\alpha_s\, \left [j^{\rm ph}_{{}_Z}(\zeta )\right ] = j^{\rm
ph}_{{}_Z}(\zeta + s)\, ,\qquad
\alpha_s\, \left [A^{\rm ph}(x ;\xi)\right ] =
\e^{is(\mu_{{}_R}\sigma + \mu_{{}_L}\tau )}\, A^{\rm ph}((x^0+s, x^1);\xi)
\, ,
\end{equation}
one can verify that the physical correlation functions (\ref{phd}) obey the
KMS condition relative to $\alpha_s$ with (inverse) temperature $\beta $. In
particular,
\bea
&&\langle A^{\rm ph}(x_1;\xi_1)\cdots A^{\rm ph}(x_m;\xi_m)
\alpha_{s+i\beta} \left [A^{\rm ph}(x_{m+1};\xi_{m+1})\cdots
A^{\rm ph}(x_n;\xi_n )\right ]
\rangle_{\mu_{{}_L},\, \mu_{{}_R}}^{{\rm ph},\, \beta } = \nb \\
&&\langle \alpha_s \left [A^{\rm ph}(x_{m+1};\xi_{m+1})\cdots
A^{\rm ph}(x_n;\xi_n )\right ]
A^{\rm ph}(x_1;\xi_1)\cdots A^{\rm ph}(x_m;\xi_m)
 \rangle_{\mu_{{}_L},\, \mu_{{}_R}}^{{\rm ph},\, \beta }
\, ,
\ena
for any $1\leq m \leq n$. This result is fundamental for the physical 
interpretation of our construction. 

At zero temperature (i. e. in the limit $\beta \to \infty $), the
anyon correlation functions are in addition invariant with respect to
dilatations
\beeq
A^{\rm ph}(x;\xi ) \longmapsto \varrho^{\frac{1}{ 2}(\sigma^2 + \tau^2)}\,
A^{\rm ph}(x;\xi )
\, , \qquad \varrho > 0 \, ,
\label{dil}
\end{equation}
and Lorentz transformations
\beeq
A^{\rm ph}(x;\xi ) \longmapsto
\e^{\frac{1}{ 2}(\sigma^2 - \tau^2)\chi } \, A^{\rm ph}(\Lambda (\chi ) x;\xi
) \, ,
\qquad \chi \in \R \, ,
\label{boost}
\end{equation}
where $\Lambda (\chi )$ is the 1+1 dimensional Lorentz boost
with parameter $\chi $. From
(\ref{dil},\ref{boost}) we deduce the dimension
\beeq
d(\xi ) \equiv \frac{1}{ 2}(\sigma^2 + \tau^2) \, ,
\label{dim}
\end{equation}
and the Lorentz spin
\beeq
l(\xi ) \equiv \frac{1}{2}(\sigma^2 - \tau^2) =
\frac{1}{2}\, \vartheta (\xi ) \,
\label{spin}
\end{equation}
of $A^{\rm ph}(x;\xi )$. In spite of the fact that transformations
(\ref{dil},\ref{boost}) do not define exact 
symmetries at finite temperature, we would like to mention that 
there exist an alternative conformal field theory approach to 
bosonization (see e.g. \cite{DCM} and references therein). 

Let us collect now some of the basic properties of
$\{ A^{\rm ph}(x;\xi )\, , j^{\rm ph}_{{}_Z}(\zeta )\, :\, \xi \in \Xi\}$, 
which follow from their correlation functions. Combining 
(\ref{acon}) and (\ref{ren}), one gets
\beeq
A^{{\rm ph}\, *}(x;\xi ) =  A^{\rm ph}(x;-\xi ) \quad .
\end{equation}
The exchange matrix and the statistical parameter of the physical anyon
fields are given
by (\ref{rdef}) and (\ref{stat}) respectively, whereas (\ref{jac}) implies
\beeq
[j^{\rm ph}_{{}_Z}(\zeta ) \, , \, A^{\rm ph}(x;\xi) ] =
2\xi^Z \delta (\zeta - x^Z) A^{\rm ph}(x;\xi) \, ,
\end{equation}
where the notation (\ref{not}) has been used. The translation of eq. (\ref{da}) 
in terms of physical fields reads 
\beeq
i\pi \xi^Z\, \vdots \, j^{\rm ph}_{{}_Z}(x^Z)A^{\rm ph}(x;\xi )\, \vdots =
\frac{\der}{ \der x^Z } A^{\rm ph}(x;\xi )
\, .
\label{npr}
\end{equation}
where the normal product $\vdots \, \cdots \, \vdots $ is defined by 
\beeq
\vdots \, j^{\rm ph}_{{}_Z}(x^Z)A^{\rm ph}(x;\xi )\, \vdots \equiv 
\lim_{y\to x} \left \{ j^{\rm ph}_{{}_Z}(y^Z)A^{\rm ph}(x;\xi ) + 
\frac{i\xi^Z}{ \beta} \coth \left [\frac{\pi}{  \beta }(y-x)^Z - i\epsilon \right ] 
A^{\rm ph}(x;\xi ) \right \} \, . 
\label{ja} 
\end{equation} 
The counterpart of eq. (\ref{lim}) is
\beeq
\Theta_{{}_Z}^{\rm ph} (\zeta) =
\frac{1}{ 4}\lim_{\zeta'\to\zeta}\left[\pi \jz^{\rm ph} (\zeta') \jz^{\rm ph}
(\zeta)
+ \frac{1}{ \pi (\zeta'-\zeta-i\epsilon)^2}\right] \, ,
\label{tlim}
\end{equation}
and using the physical correlation functions one finds:
\bea
&&[\Theta_{{}_Z}^{\rm ph} (\zeta)\, ,\, A^{\rm ph}(x;\xi) ] =
-i\delta (\zeta - x^Z)\, \frac{\der}{  \der x^Z } A^{\rm ph}(x;\xi)  \, , 
\label{ta} \\ 
&&
[\Theta_{{}_Z}^{\rm ph} (\zeta_1)\, ,\, \Theta_{{}_Z}^{\rm ph} (\zeta_2)] =
\delta^\prime (\zeta_{12})[\Theta_{{}_Z}^{\rm ph} (\zeta_1) +
\Theta_{{}_Z}^{\rm ph} (\zeta_2)]
-i \frac{1}{ 24\pi } \delta^{\prime \prime \prime}(\zeta_{12}) \, .
\label{tt}
\ena
Eqs. (\ref{ta},\ref{tt}) provide a consistency check on the physical
correlation functions (\ref{phd}) and (\ref{tlim}). Moreover, 
the $\delta^{\prime \prime \prime}\, $-contribution in the right hand side 
of eq. (\ref{tt}) fixes the value of the central charge. As it should be 
expected on general grounds, the latter is $\beta $- and 
$\mu_{{}_Z}$-independent. 

{}Finally, from eq. (\ref{xxx}) one gets
\bea
&&C^{\rm ph}(x,\eta;\xi) \equiv
\frac{1}{ 2}\left [A^{{\rm ph}\, *}(x + \eta ;\xi)\, A^{\rm ph}(x;\xi) -
A^{\rm ph}(x;\xi)\, A^{{\rm ph}\, *}(x-\eta ;\xi)\right ] = \nb \\
&&i\pi \, (2\pi)^{-d(\xi )}\, \e^{\pi [\sigma^2 u(\eta^-) + \tau^2 u(\eta^+)]}
\, \left [\sigma \eta^- j^{\rm ph}_{{}_R}(x^-) +
\tau \eta^+ j^{\rm ph}_{{}_L}(x^+) + O(\eta^+) + O(\eta^-)\right ]\, .
\label{cph}
\ena
This identity suggests that $A^{\rm ph}$ and $j_{{}_Z}^{\rm ph}$ are not
independent and that one can recover $j_{{}_Z}^{\rm ph}$ from $A^{\rm ph}$
by point-splitting. For this purpose we assume that $\eta $ is a
space-like vector ($\eta^2 = \eta^+ \eta^- < 0$) and observe that
the behavior of $C^{\rm ph}(x,\eta;\xi)$ when $\eta \to 0$ is
direction dependent. Setting
\beeq
Z_{{}_Z}(\eta ;\xi ) = \frac{(2\pi)^{d(\xi)}}{ i\pi \eta^Z}
\e^{-\pi [\sigma^2 u(\eta^-) + \tau^2 u(\eta^+)]}
\, ,
\end{equation}
one gets from (\ref{cph})
\bea
&&\sigma j_{{}_R}(x^-) = \lim_{\eta^1 \downarrow 0} \left [\, \lim_{\eta^0
\downarrow -\eta^1}
\, Z_{{}_R} (\eta ;\xi )\, C^{\rm ph} (x, \eta;\xi ) \right ] \quad ,
\label{lim1}
\\
&&
\tau j_{{}_L}(x^+) = \lim_{\eta^1 \downarrow 0} \left [\, \lim_{\eta^0
\uparrow \eta^1}
\, Z_{{}_L} (\eta ;\xi )\, C^{\rm ph} (x, \eta;\xi ) \right ] \quad .
\label{lim2}
\ena
We stress that the two limits in the right hand side of (\ref{lim1}) and
(\ref{lim2})
do not commute and must be taken in the prescribed order. One constructs in this
way $j^{\rm ph}_{{}_R}$ ($j^{\rm ph}_{{}_L}$) from $A^{\rm ph} (x;\xi )$,
provided that
$\sigma \not= 0$ ($\tau \not=0$). We would like to mention that the inverse
operation is also possible. $A^{\rm ph} (x;\xi )$ can be expressed in terms of
$j_{{}_Z}$, which allows to develop a framework for bosonization, totally based 
on the algebra of chiral currents. We refer for the zero temperature case
to \cite{St} and \cite{DFZ}.

Summarizing, with any set $\Xi $ a thermal anyon representation 
$\T(\Xi ; \mu_{{}_L}, \mu_{{}_R})$ is associated, which is defined by 
the expectation values (\ref{phd}). Let
us consider the set 
$\Xi = \{\xi,\, -\xi,\, \xi^\prime,\, -\xi^\prime\}$ with
\beeq
\xi = (\sigma,\, \tau ) \, , \qquad \xi^\prime = (\tau,\, \sigma )\, , \qquad
\vartheta (\xi ) = \sigma^2 - \tau^2 \not= 0 \, .
\label{par}
\end{equation}
The corresponding representation will be denoted
by $\T(\sigma , \tau ; \mu_{{}_L}, \mu_{{}_R})$ and is essential
in the bosonization of fermions.

In what follows, we will use exclusively the physical anyon representations
constructed above, omitting for sake of notation 
simplicity the apex ``ph". One should always
keep in mind however that $\{ A(x;\xi )\, , j_{{}_Z}(\zeta )\, :\, \xi \in
\Xi\}$
and $\{ A^{\rm ph}(x;\xi )\, , j^{\rm ph}_{{}_Z}(\zeta )\, :\, \xi \in \Xi\}$
are different quantum fields.

\subsection{\bf Momentum distribution and anyon condensation}

In order to get a deeper insight into the physical properties of the field
$A(x;\xi )$, it is instructive to derive the relative momentum 
distribution. For this purpose we consider the Fourier transform 
\beeq
{\widehat W}^\beta (\omega , k;\xi ) = \int_{-\infty }^\infty 
d^2x \, 
\e^{i\omega x^0 - ikx^1}  W^\beta (x ;\xi )  
\label{ft}  
\end{equation}
of the two-point function 
\beeq
W^\beta (x_{12};\xi ) \equiv \langle A^*(x_1;\xi)A(x_2;\xi) 
\rangle_{\mu_L,\, \mu_R}^\beta \, . 
\end{equation}
Employing eqs. (\ref{ac},\ref{phd}) one easily derives 
\beeq 
{\widehat W}^\beta (\omega , k;\xi ) = \frac {1}{2}\,  
\varrho \left (\frac{\omega }{2} + \frac{k}{2} + 
\mu_{{}_R} \sigma ;\, \sigma^2 , \beta \right ) 
\varrho \left (\frac{\omega }{2} - \frac{k}{2} + 
\mu_{{}_L} \tau ;\, \tau^2 , \beta \right ) \, , 
\label{roro} 
\end{equation}
where 
\beeq
\varrho (k ;\alpha  ,\, \beta ) = \int_{-\infty}^\infty d\zeta \,e^{ik\zeta }
\,\left [2i\beta \, \sinh \left (\frac{\pi}{ \beta }\zeta -i\epsilon \right )
\right ]^{-\alpha }\, , \qquad \alpha \geq 0\, .
\label{fou}
\end{equation}
\begin{figure}[h]
\begin{picture}(10,10)
%put(200,-215){$ k$}
\put(385, -170){$\alpha$}
\end{picture}
\vskip -2 truecm 
\centerline{\hskip 3.5truecm \epsfig{file=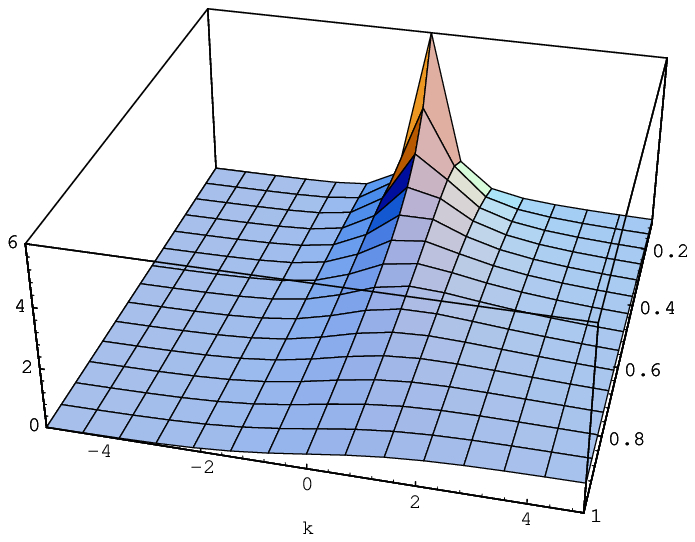,height=24cm,width=20cm}}
\vskip -13 truecm 
\centerline{{\bf Figure 1}: The momentum distribution $\varrho (k; \alpha , \beta=1$)}
\vskip 0.5 truecm 
\end{figure}
The evaluation of the integral (\ref{fou}) gives
\beeq 
\varrho (k ;\alpha ,\, \beta ) = \left \{ \begin{array}{cc}(\beta )^{1-\alpha}
\, \e^{\frac{1}{ 2}\beta k }\, \Big| \Gamma \left (\frac{1}{  2}\alpha +
\frac{i}{ 2\pi }\beta k \right ) \Big|^2 [2\pi \Gamma (\alpha )]^{-1}& 
\mbox{for } \alpha > 0\, , \\ 2 \pi \, \delta (k) & \mbox{for } \alpha = 0 \, . 
\end{array} \right .
\label{dis}
\end{equation}

Since the contributions of the left- and right-moving modes factorize, 
it is convenient to consider first the particular cases 
\beeq 
{\widehat W}^\beta (\omega , k;(\sigma ,0)) = 
2\pi \delta (\omega - k) 
\varrho \left (k + \mu_{{}_R} \sigma ;\, \sigma^2 , \beta \right ) \, , 
\label{rro} 
\end{equation}
and 
\beeq 
{\widehat W}^\beta (\omega , k;(0,\tau )) = 
2\pi \delta (\omega + k) 
\varrho \left (-k + \mu_{{}_L} \tau ;\, \tau^2 , \beta \right ) \, . 
\label{lro} 
\end{equation}
Eqs. (\ref{rro},\ref{lro}) have a simple physical interpretation: 
the $\delta $-factors fix the dispersion 
relations, whereas the $\varrho $-factors give the momentum distributions. 
Eq. (\ref{rro}) for $\sigma = \pm 1$ and eq. (\ref{lro}) for $\tau = \pm 1$ 
provide useful checks. From the exchange relation (\ref{exc}) we already know  
that the fields $A(x;(\pm 1,0))$ and $A(x;(0,\pm 1))$ have Fermi statistics. 
In the next section we will show that they are actually canonical right and 
left chiral fermions. In agreement with this fact, setting 
$\alpha =1$ in eq. (\ref{dis}), we get from (\ref{rro},\ref{lro}) 
\beeq 
{\widehat W}^\beta (\omega , k;(\pm 1,0)) = 
2\pi \delta (\omega - k) \, \frac{1}{ 1+\e^{-\beta (k\, \pm \, \mu_{{}_R})}}\, , 
\end{equation}
\label{rdd}
\beeq
{\widehat W}^\beta (\omega , k;(0,\pm 1)) = 
2\pi \delta (\omega + k)\, \frac{1}{ 1+\e^{\beta (k\, \mp \,\mu_{{}_L})}} \, .
\label{ldd}
\end{equation}
As expected, the familiar Fermi distribution at finite
temperature and chemical potential is recovered. 
The corresponding Fermi momenta are 
\beeq
k_{{}_F} = \mp \, \mu_{{}_R} \, , \qquad k_{{}_F} = \pm \, \mu_{{}_L} \, .
\end{equation} 
\begin{figure}[h]
\begin{picture}(10,10)
\put(247,-213){$ k$}
\end{picture}
\vskip -2 truecm 
\centerline{\hskip 3.5truecm \epsfig{file=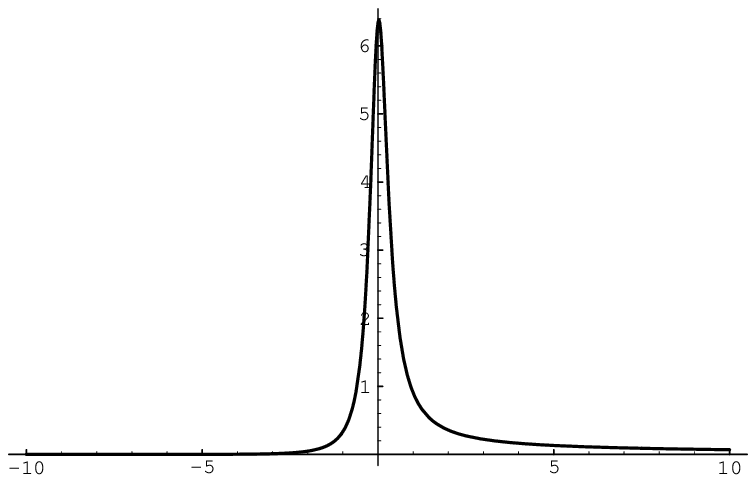,height=24cm,width=20cm}}
\vskip -14 truecm 
\centerline{{\bf Figure 2}: Condensation-like behavior of 
$\varrho (k; \alpha=10^{-1}, \beta=1$)}
\vskip 0.5 truecm 
\end{figure}
\begin{figure}
\begin{picture}(10,10)
\put(247,-173){$ k$}
\end{picture}
\vskip -2 truecm 
\centerline{\hskip 3.5truecm \epsfig{file=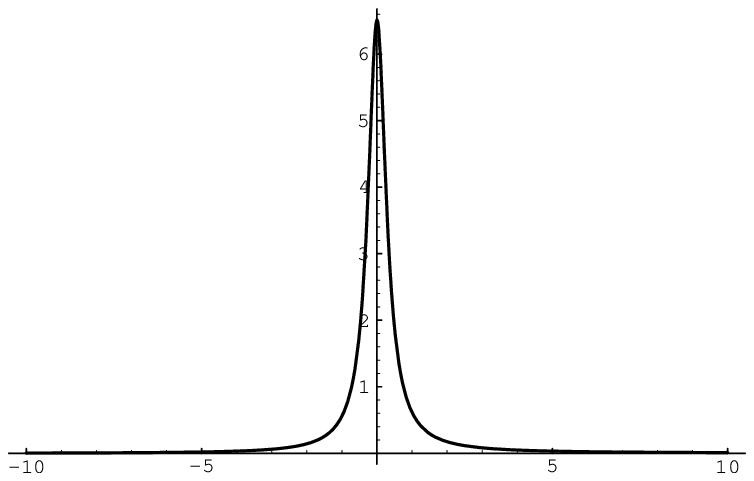,height=20cm,width=20cm}}
\vskip -11.5 truecm 
\centerline{{\bf Figure 3}: The momentum distribution $\varrho (k; \alpha = 
10^{-2}, \beta=10^{-1})$}
\vskip 0.5 truecm 
\end{figure}
 
Turning back to the general expression (\ref{dis}), we see that for $\alpha >0$ 
the distribution $\varrho $ is a smooth positive function, whose asymptotic 
behavior is encoded in
\bea
&& \varrho (k;\alpha ,\beta ) \sim \frac{1}{  \Gamma (\alpha )} 
\left (\frac{k}{  2\pi}\right )^{\alpha -1} \, , \, \, \, \, 
\qquad k\to \infty \, , 
\label{kp}\\
&&\varrho (k;\alpha ,\beta ) \sim \frac{e^{\beta k}}
{ \Gamma (\alpha )}\left (-\frac{k}{ 2\pi}\right )^{\alpha -1} , 
\qquad k\to -\infty \, . 
\label{km}
\ena

One can get a general idea about $\varrho $ from 3D plot in figure 1. 
In the range $\alpha \geq 1$, $\varrho $ is monotonically increasing on the whole 
line $k \in \R$. When $0 < \alpha < 1$, $\varrho $ increases monotonically 
for $k\leq 0$ and, according to eqs. (\ref{kp},\ref{km}), admits at least one 
local maximum for $k > 0$. Let us denote the position of the first one (when $k$ 
moves from $0$ to $\infty $) by $k_{{}_C}(\alpha, \beta )$. We have numerical 
evidence that this maximum is unique, but we have not an analytic 
proof of this statement. 
The plots of $\varrho $ (see figs. 1-2) indicate 
an interesting condensation-like behavior around $k_{{}_C}$. The phenomenon 
is clearly marked for small values of the temperature and/or of the parameter 
$\alpha $ in the domain $0 < \alpha < 1$. We find it quite 
remarkable that for any fixed temperature, one can achieve an arbitrary sharp 
anyon condensation, taking a sufficiently small $\alpha $.
See in this respect figure 3, where the momentum distribution is plotted for 
$\beta = 10^{-1}$. 

Concerning the behavior of ${\widehat W}^\beta (\omega , k;\xi )$ 
(see eq. (\ref{roro})) when both $\sigma \not=0$ and $\tau \not= 0$, 
the above analysis implies condensation at 
\beeq 
\omega = k_{{}_C}(\sigma^2, \beta) + k_{{}_C}(\tau^2, \beta) - 
\mu_{{}_R}\sigma - \mu_{{}_L}\tau \, , 
\end{equation}
\beeq 
k = k_{{}_C}(\sigma^2, \beta) - k_{{}_C}(\tau^2, \beta) - 
\mu_{{}_R}\sigma + \mu_{{}_L}\tau \, , 
\end{equation}
provided that $0<\sigma^2 <1$ and $0<\tau^2 <1$ 

It is worth mentioning that the condensation effect we discovered, does not contradict 
the Hohenberg-Mermin-Wagner (HMW) theorem about the absence of condensation in 
one space dimension. The point is that under very general conditions, any 
anyonic statistics can be equivalently described by a suitable exchange 
interaction with two- and three-body potentials, determined \cite{LiMi} 
by the exchange factor (\ref{rdef}). When these potentials are confining, 
some assumptions of the HMW theorem are violated and condensation may occur 
\cite{BK}, \cite{No} even in one dimension. 

In conclusion, we have shown that the right 
(left) moving modes of the anyon field $A(x;\xi )$ condensate 
in the range $0<\sigma^2 < 1$ ($0<\tau^2 < 1$). Some 
applications of this phenomenon are currently under investigation \cite{LMP}.

\section{\bf A particular case - fermions}

An useful testing ground for the bosonization scheme proposed above is 
provided by the free fermion field. The possibility to express the 
latter in terms of bosonic fields,
was discovered long ago by Jordan and Wigner \cite{JW}. Using the general
results of the previous section, we will discuss below
some aspects of this phenomenon at finite temperature and
density. Let us start by the 1+1 dimensional massless free Dirac equation
\beeq
\gamma^\nu \der_\nu \, \psi (x) = 0 \quad ,
\label{dir}
\end{equation}
where
\beeq
\psi (x)= \left( \begin{array}{c} \psi_1(x) \\ \psi_2(x) \end{array} \right)
\quad, \qquad
\gamma^0=  \left( \begin{array}{cc} 0 & 1 \\ 1 & 0 \end{array} \right) \quad,
\qquad
\gamma^1=  \left( \begin{array}{cc} 0 & -1 \\ 1 &   0 \end{array} \right)
\quad .
\end{equation}
One easily verifies that
\beeq
\psi_1(x) = \int_0^\infty \mis \left [d^*(k)\, \e^{ikx^-} + b(k)\,
\e^{-ikx^-} \right ]
\equiv \spr (x^-) \quad ,
\label{f1}
\end{equation}
\beeq
\psi_2(x) = \int^0_{-\infty }\mis \left [ b(k)\, \e^{ikx^+} - d^*(k)\,
\e^{-ikx^+} \right ]
\equiv \spl (x^+)
\quad ,
\label{f2}
\end{equation}
obey eq. (\ref{dir}). Here $\{b^*(k),\, b(k),\, d^*(k),\, d(k) \}$ are the
generators
of an associative algebra $\A_+$ (with involution ${}^*$ and identity
element $\1$),
which are assumed to satisfy the anti-commutation relations
\bea
&&\{b(k)\, ,\, b(p)\} = \{b^*(k)\, ,\, b^*(p)\} =
\{d(k)\, ,\, d(p)\} = \{d^*(k)\, ,\, d^*(p)\} = 0
\quad , \nb \\
&&\{b(k)\, ,\, b^*(p)\} = \{d(k)\, ,\, d^*(p)\} = 2\pi \delta (k-p)\, \1
\quad .
\label{car}
\ena
Combining eqs. (\ref{f1}-\ref{f2}) and (\ref{car}), one finds
\beeq
\{\psi (x_1)\, ,\, \psi (x_2)\} = 0 \, , \qquad
\{\psi (x_1)\, ,\, \psi^* (x_2)\} =
\left( \begin{array}{cc} \delta (x_{12}^-)  & 0 \\ 0 & \delta (x_{12}^+)
\end{array} \right) \, ,
\label{cfr}
\end{equation}
which in particular imply the equal time canonical anti-commutation
relations.

As a consequence of (\ref{dir}), both the vector and the axial currents
\beeq
j_\mu (x) =\, \, :\overline \psi \gamma_\mu \psi : (x) \, ,
\qquad j_\mu^5 (x) =\, \, :\overline \psi \gamma_\mu \gamma^5 \psi :(x) \, ,
\qquad \gamma^5 \equiv \gamma^0\gamma^1 \, ,
\label{vac}
\end{equation}
are conserved. Here $\overline \psi \equiv \psi^\ast \gamma^0 $ and 
the normal product is
relative to the creators $\{b^*(k),\, d^*(k)\}$ and the annihilators
$\{b(k),\, d(k)\}$. The duality relation
\beeq
j_\mu^5 (x) = \varepsilon_{\mu \nu}\, j^\nu (x) \quad ,
\label{dual1}
\end{equation}
holds because of the identity
$\gamma_\mu \gamma^5 = \varepsilon_{\mu \nu}\gamma^\nu $. One has
\bea
&&[j_0(x)\, ,\, \psi (y)]\vert_{x^0=y^0} = - \delta (x^1-y^1) \psi(y)\, ,
\label{jpsi} \\
&&[j^5_0(x)\, ,\, \psi (y)]\vert_{x^0=y^0} = - \delta (x^1-y^1) \gamma^5\,
\psi(y) \,  .
\label{cucr}
\ena
The conserved charges relative to $j_\mu $ and $j_\mu^5$ read
\beeq
Q = \int_{-\infty }^\infty \mis [b^*(k)b(k) - d^*(k)d(k)]\, , \qquad
Q_{{}_5} = \int_{-\infty }^\infty \mis \varepsilon (k) [b^*(k)b(k) -
d^*(k)d(k)] \, .
\end{equation}

Let us consider now the thermal representation $\T_+$ of $\A_+$, whose
cyclic vector
is the Gibbs equilibrium state associated with
\beeq
K_{{}_F} = H_{{}_F} - \mu Q - \mu_{{}_5} Q_{{}_5} \, \qquad \mu,\,
\mu_{{}_5} \in \R \,
\end{equation}
where
\beeq
H_{{}_F} = \int_{-\infty }^\infty \mis |k|\, [b^*(k)b(k) + d^*(k)d(k)] \quad ,
\end{equation}
is the Hamiltonian of the free Dirac field. We stress the presence of
the chemical potential $\mu_{{}_5}$, associated with the chiral symmetry.
The non-vanishing two-point expectation values in this equilibrium state are:
\beeq
\langle b^*(p)b(q)\rangle_{\mu,\, \mu_{{}_5}}^\beta =
\frac{\e^{-\beta [|p|-\mu - \varepsilon (p)\mu_{{}_5}]}
}{ 1+\e^{-\beta [|p|-\mu - \varepsilon (p)\mu_{{}_5}]}}
\, 2\pi \delta (p-q) \, ,
\label{bb}
\end{equation}
\beeq
\langle b(p)b^*(q)\rangle_{\mu,\, \mu_{{}_5}}^\beta =
\frac{1}{ 1+\e^{-\beta [|p|-\mu - \varepsilon (p)\mu_{{}_5}]}}
\, 2\pi \delta (p-q) \, ,
\end{equation}
\beeq
\langle d^*(p)d(q)\rangle_{\mu,\, \mu_{{}_5}}^\beta =
\frac{\e^{-\beta [|p|+\mu + \varepsilon (p)\mu_{{}_5}]}
}{ 1+\e^{-\beta [|p|+\mu + \varepsilon (p)\mu_{{}_5}]}}
\, 2\pi \delta (p-q) \, ,
\end{equation}
\beeq
\langle d(p)d^*(q)\rangle_{\mu,\, \mu_{{}_5}}^\beta =
\frac{1}{ 1+\e^{-\beta [|p|+\mu + \varepsilon (p)\mu_{{}_5}]}}
\, 2\pi \delta (p-q) \, .
\label{dd}
\end{equation}
Like in the bosonic case, by using
$\langle \1 \rangle_{\mu,\, \mu_{{}_5}}^\beta = 1$ a generic
$n$-point function can be expressed in terms of (\ref{bb}-\ref{dd}). The
correlation functions of $\psi $ are therefore fully determined. The ones
involving
$\psi^*_a$ and $\psi_a$ with the same value of the index $a$, can be reduced
by means of eqs. (\ref{cfr}) to
\bea
&&\langle \psi_1^* (x_1)\cdots \psi_1^* (x_m)\psi_1 (y_n)
\cdots \psi_1 (y_1)\rangle_{\mu,\, \mu_{{}_5}}^\beta =
\delta_{mn}\, {\rm det}\langle \psi_1^* (x_i)\psi_1 (y_j)\rangle_{\mu,\,
\mu_{{}_5}}^\beta = \nb \\
&& \delta_{mn} \exp \left [i(\mu +\mu_{{}_5})\sum_{i,j=1}^n (x_i-y_j)^-\right ]
\, {\rm det} \frac{1}{ 2i\beta \sinh \left [\frac{\pi}{ \beta }(x_i-y_j)^-
-i\epsilon \right ]} \, , \label{x} \\
\label{f}
&& \langle \psi_2^* (x_1)\cdots \psi_2^* (x_m)\psi_2 (y_1)
\cdots \psi_2 (y_n)\rangle_{\mu,\, \mu_{{}_5}}^\beta =
\delta_{mn}\, {\rm det}\langle \psi_2^* (x_i)\psi_2 (y_j)\rangle_{\mu,\,
\mu_{{}_5}}^\beta = \nb \\
&&\delta_{mn} \exp \left [i(\mu -\mu_{{}_5})\sum_{i,j=1}^n (x_i-y_j)^+\right ]
\, {\rm det} \frac{1}{ 2i\beta \sinh \left [\frac{\pi}{  \beta }(x_i-y_j)^+
-i\epsilon \right ]}
\, .
\label{xx}
\ena
The expectation values involving both $\psi_1$ and $\psi_2$ factorize in
the product of
two correlators depending only on $\psi_1$ and $\psi_2$ respectively. For
the currents (\ref{vac}), one gets
\beeq
\langle j_0 (x) \rangle_{\mu,\, \mu_{{}_5}}^\beta = \frac{\mu}{ \pi} \, ,
\qquad
\langle j_0^5 (x) \rangle_{\mu,\, \mu_{{}_5}}^\beta = \frac{\mu_{{}_5}}{
\pi} \, .
\label{exc1}
\end{equation}
It follows from eq. (\ref{dual1}) 
\beeq
\langle j_1 (x) \rangle_{\mu,\, \mu_{{}_5}}^\beta = -\frac{\mu_{{}_5}}{ \pi}
\, , \label{exc2} 
\end{equation}
which clarifies the physical meaning of the chemical potential $\mu_{{}_5}$.
It gives rise to a persistent current at finite temperature. We will discuss
later on this interesting phenomenon in the context of the Thirring model.

Let us show now that the free fermion field considered above, 
has an equivalent bosonized description. For this purpose we take 
the representation $\T(\sigma, 0;\mu_{{}_L},\, \mu_{{}_R})$ and set
\beeq
\Psi (x;\sigma ) = \left( \begin{array}{c} \Psi_1(x;\sigma ) \\
\Psi_2(x;\sigma ) \end{array} \right)
\equiv  \left( \begin{array}{c} A(x;(\sigma, 0)) \\ A(x;(0, \sigma))
\end{array} \right) \, .
\label{bos}
\end{equation}
Since $\Psi_1 $ and $\Psi_2$ depend on $x^-$ and $x^+$ respectively,
the field $\Psi $ obviously satisfies eq. (\ref{dir}). According to eq.
(\ref{stat}),
the corresponding statistical parameter is
\beeq
\vartheta = \sigma^2 \, , \qquad \sigma \not= 0 \, ,
\end{equation}
showing that, in general, $\Psi $ obeys anyon statistics. The existence of
anyonic
solutions of the free Dirac equation (\ref{dir}) is not surprising.
Let us recall
in this respect that there is no genuine notion of spin in 1+1 dimensions, because
the rotation group is trivial. Being associated with the one-parameter group of
boost transformations, which is non-compact, the Lorentz spin $l(\xi )$
(see eq. (\ref{spin})) may take any real value.

At this point we need the correlation functions of $\Psi $. From
eqs.(\ref{ac},\ref{phd}), one gets
\bea
&&\langle \Psi_1^* (x_1;\sigma )\cdots \Psi_1^* (x_m;\sigma )
\Psi_1 (y_n;\sigma )
\cdots \Psi_1 (y_1;\sigma )\rangle_{\mu_{{}_L},\, \mu_{{}_R}}^\beta =
\delta_{mn} \exp \left [i\mu_{{}_R}\sigma \sum_{i,j=1}^n (x_i-y_j)^-\right ]
\nb \\
&&
\frac{\prod_{{}_{\stackrel{i,j=1}{ i<j}}}^n
\left [2i\beta \sinh \left (\frac{\pi}{ \beta }x_{ij}^- -i\epsilon \right
)\right ]^{\sigma^2}
\prod_{{}_{\stackrel{i,j=1}{ i>j}}}^n
\left [2i\beta \sinh \left (\frac{\pi}{ \beta }y_{ij}^- -i\epsilon \right
)\right ]^{\sigma^2}}
{ \prod_{{}_{i,j=1}}^n
\left \{ 2i\beta \sinh \left [\frac{\pi}{ \beta }(x_i-y_j)^- -i\epsilon
\right ]\right \}^{\sigma^2}}
\, , \label{cf} \\
&&
\langle \Psi_2^* (x_1;\sigma )\cdots \Psi_2^* (x_m;\sigma )\Psi_2 (y_n;\sigma )
\cdots \Psi_2 (y_1;\sigma )\rangle_{\mu_{{}_L},\, \mu_{{}_R}}^\beta =
\delta_{mn} \exp \left [i\mu_{{}_L}\sigma \sum_{i,j=1}^n (x_i-y_j)^+\right ]
\nb \\
&&
\frac{\prod_{{}_{\stackrel{i,j=1}{ i<j}}}^n
\left [2i\beta \sinh \left (\frac{\pi}{ \beta }x_{ij}^+ -i\epsilon \right
)\right ]^{\sigma^2}
\prod_{{}_{\stackrel{i,j=1}{ i>j}}}^n
\left [2i\beta \sinh \left (\frac{\pi}{ \beta }y_{ij}^+ -i\epsilon \right
)\right ]^{\sigma^2}}
{ \prod_{{}_{i,j=1}}^n
\left \{ 2i\beta \sinh \left [\frac{\pi}{ \beta }(x_i-y_j)^+ -i\epsilon
\right ]\right \}^{\sigma^2}}
\, .
\label{cff}
\ena
Since the mixed $\Psi $-correlators factorize like in the $\psi $-case, it
is enough
to compare eqs. (\ref{cf},\ref{cff}) with eqs. (\ref{x},\ref{xx}). For
this purpose we observe that the
$i\epsilon$-prescription in the numerator of (\ref{cf}) and (\ref{cff}) is
actually superfluous and the identity
\beeq
{\rm det}\, \frac{1}{ 2i\beta \sinh \left [\frac{\pi}{ \beta }(x_i-y_j)
-i\epsilon \right ]} =
\frac{\prod_{{}_{\stackrel{i,j=1}{ i<j}}}^n 2i\beta \sinh \left (\frac{\pi}{ \beta
}x_{ij}
\right )
\, \, \prod_{{}_{\stackrel{i,j=1}{ i>j}}}^n 2i\beta \sinh \left (\frac{\pi}{ \beta
}y_{ij} \right )
}{ \prod_{{}_{i,j=1}}^n
2i\beta \sinh \left [\frac{\pi}{ \beta }(x_i-y_j) -i\epsilon \right ]}
\, .
\label{trick}
\end{equation}
holds. The proof of Eq. (\ref{trick}) is given in the appendix. 
Eq. (\ref{trick}) generalizes
\beeq
{\rm det}\, \frac{1}{ (x_i-y_j -i\epsilon )} =
\frac{\prod_{{}_{\stackrel{i,j=1}{ i<j}}}^n (x_{ij})\, \, \prod_{{}_{\stackrel
{i,j=1}{i>j}}}^n(y_{ij})}{\prod_{{}_{i,j=1}}^n
(x_i-y_j -i\epsilon ) }
\, \, ,
\label{trick1}
\end{equation}
which makes possible the bosonization at zero temperature.

In view of (\ref{trick}), we conclude that the correlation functions of
$\psi (x)$ and $\Psi (x;\sigma)$ coincide, provided that
\beeq
\sigma^2 = 1 \quad ,
\label{cond1}
\end{equation}
and
\beeq
\mu_{{}_R} = \frac{1}{ \sigma }(\mu + \mu_{{}_5}) \, , \qquad
\mu_{{}_L} = \frac{1}{ \sigma }(\mu - \mu_{{}_5}) \, .
\label{cond2}
\end{equation}
In other words, the representations
$\T(1, 0;\mu + \mu_{{}_5},\, \mu - \mu_{{}_5})$ and
$\T(-1, 0;-\mu - \mu_{{}_5},\, -\mu + \mu_{{}_5})$ are equivalent and
yield the bosonized version of the field $\psi $. For
\beeq
\sigma^2 = 2n+1 \, , \qquad n\geq 1 \,
\end{equation}
$\Psi (x;\sigma )$ still possess Fermi statistics, but is not canonical. 
According to the results of the previous section, the momentum distributions 
of both $\Psi_1 (x;\sigma )$ and $\Psi_2 (x;\sigma )$ signal condensation 
in the range $0< \sigma^2 < 1$. 

{}For completing the Fermi-Bose correspondence, we have to find in
$\T(\sigma, 0;\mu_{{}_L},\, \mu_{{}_R})$ the counterparts $J_\mu $ and
$J^5_\mu $ of the fermionic currents (\ref{vac}). Let us define
\bea
&&J_0(x) = -J^5_1 (x) = -\frac{1}{ 2\sigma } [j_{{}R}(x^-) + j_{{}_L}(x^+)]
\, ,  \label{ct} \\
&&J_1(x) = -J^5_0 (x) = \frac{1}{ 2\sigma } [j_{{}R}(x^-) - j_{{}_L}(x^+)]
\, .
\label{cs}
\ena
These currents are conserved by construction. Moreover, eq. (\ref{jac})
implies
\bea
&&[J_0(x)\, ,\, \Psi (y)]\vert_{x^0=y^0} = - \delta (x^1-y^1) \Psi(y)
\quad , \label{fbc1} \\
&&[J^5_0(x)\, ,\, \Psi (y)] \vert_{x^0=y^0} = - \delta (x^1-y^1) \gamma^5\,
\Psi(y) \quad ,
\label{fbc2}
\ena
which precisely reproduce eqs. (\ref{jpsi},\ref{cucr}). Notice that the
normalization in (\ref{ct},\ref{cs}) is uniquely fixed by requiring
(\ref{fbc1},\ref{fbc2}). Using (\ref{exj}), one finds
\beeq
\langle J_0 (x) \rangle_{\mu_{{}_L},\, \mu_{{}_R}}^\beta =
\frac{1}{ 2\pi \sigma }(\mu_{{}_R} + \mu_{{}_L}) \, , \qquad
\langle J_0^5 (x) \rangle_{\mu_{{}_L},\, \mu_{{}_R}}^\beta =
\frac{1}{ 2\pi \sigma }(\mu_{{}_R} - \mu_{{}_L})\, .
\label{excb}
\end{equation}
As it should be expected, (\ref{exc1}) and (\ref{excb}) are equivalent, if
(\ref{cond1},\ref{cond2}) are satisfied.

\section{\bf The Thirring model}

The Thirring model \cite{Th} has been extensively studied in
the past. The classical equation of motion is
\beeq
i\gamma^\nu \der_\nu \Psi (x) = g\pi\, [{\overline \Psi}(x)\gamma_\nu \Psi
(x)] \gamma^\nu \Psi (x)
\, ,
\label{them}
\end{equation}
where $g \in \R$ is the coupling constant and
the factor $\pi $ has been introduced for further convenience.
Both vector and chiral currents,
\beeq
J_\nu (x) = {\overline \Psi}(x)\gamma_\nu \Psi (x) \, , \qquad
J^5_\nu (x) = {\overline \Psi}(x)\gamma_\nu \gamma^5 \Psi (x) \, ,
\label{thc}
\end{equation}
are conserved, satisfy the duality relation
\beeq
J^5_\nu (x) =\varepsilon_{\nu \mu }J^\mu (x) \, ,
\label{thdual}
\end{equation}
and have the following equal time Poisson brackets
\bea
&&\{J_0(0,x^1)\, ,\, \Psi (0,y^1)\}_{{}_{\rm P.B.}} = - \delta (x^1-y^1) \Psi
(0,y^1) \, , \\
&&\{J_0^5(0,x^1)\, ,\, \Psi (0,y^1)\}_{{}_{\rm P.B.}} = - \delta (x^1-y^1)
\gamma^5 \Psi (0,y^1) \, .
\label{poi}
\ena
Introducing
\beeq
J_+(x) = J_0(x) + J_1(x)  \, , \qquad
J_-(x) = J_0(x) - J_1(x) \, ,
\end{equation}
eq. (\ref{them}) can be rewritten in the form
\beeq
2i\der_+ \Psi_1 (x) = g\pi\, J_+(x) \Psi_1 (x)\, , \qquad
2i\der_- \Psi_2 (x) = g\pi\, J_-(x) \Psi_2 (x) \, .
\label{athe}
\end{equation}
Our problem is now to quantize this system at finite temperature and
chemical potentials
$\mu $ and $\mu^5$, associated with the charges $Q$ and $Q^5$ generated by
the currents (\ref{thc}).
More precisely, we look for operators $\{\Psi,\, J_+,\, J_- \}$ acting on a
Hilbert space and satisfying the requirements:
\begin{itemize}
\item {(a)} there exist suitable normal products $N[J_+\Psi_1] (x)$ and
$N[J_-\Psi_2] (x)$, such that the quantum versions
\beeq
2i\der_+ \Psi_1 (x) = g\pi\, N[J_+\Psi_1 ] (x)\, , \qquad
2i\der_- \Psi_2 (x) = g\pi\, N[J_-\Psi_2 ](x) \, ,
\label{qthe}
\end{equation}
of (\ref{athe}) hold;
\item {(b)} the operators
\beeq
J_0(x) = \frac{1}{ 2}[J_+(x) + J_-(x)]\, , \qquad
J_1(x) = \frac{1}{ 2}[J_+(x) - J_-(x)]\, ,
\label{coa}
\end{equation}
are the components of a conserved current $J_\nu (x)$ and
\beeq
[J_0(x)\, ,\, \Psi (y)]\vert_{x^0=y^0} = - \delta (x^1-y^1) \Psi (y) \, ;
\end{equation}
\item {(c)} the axial current $J_\nu^5 (x)$ is related to $J_\nu (x)$ via
the duality relation (\ref{thdual}) and it is also conserved;
\item {(d)} the correlation functions of $\{\Psi,\, J_+,\, J_- \}$ obey
the KMS condition relative to the automorphism
\bea
&&\alpha_s \, \Psi_1 (x) = e^{is(\mu + \mu_{{}_5})} \Psi_1 (x^0 + s, x^1) \,
, \qquad
\alpha_s \, \Psi_2 (x) = e^{is(\mu - \mu_{{}_5})} \Psi_1 (x^0 + s, x^1) \, ,
\nb \\
&& \alpha_s J_\pm (x) = J_\pm (x^0 + s, x^1) \, , \qquad s \in \R \, ,
\ena
with (inverse) temperature $\beta $.
\end{itemize}
In the previous section we have actually solved the problem defined by
(a-d) in the case
$g=0$. This solution suggests considering the representation
$\T(\sigma , \tau ; \mu_{{}_L}, \mu_{{}_R})$ (see eq. (\ref{par})) and setting 
\beeq
\Psi_1(x) \equiv A(x;\xi ) \, , \quad
\Psi_2(x) \equiv A(x;\xi^\prime ) \, ,
\end{equation}
which generalizes eq. (\ref{bos}). The next step is to identify the normal
product $N$ we are looking
for, with $\vdots \cdots \vdots $. Then, defining
\beeq
J_+(x) \equiv -\frac{1}{ \sigma + \tau }\, j_{{}_L}(x^+) \, , \qquad
J_-(x) \equiv -\frac{1}{ \sigma + \tau }\, j_{{}_R}(x^-) \, ,
\label{cthd}
\end{equation}
we see that (\ref{npr}) implies (\ref{qthe}), provided that
\beeq
2\tau (\sigma + \tau ) = g \, .
\label{coup}
\end{equation}
In addition, inserting (\ref{cthd}) in (\ref{coa}), we find that the
requirements of point (b) above
are satisfied. Let us focus on (c). The duality relation (\ref{thdual})
defines a conserved chiral
current $J_\nu^5(x)$ in terms of $J_\nu (x) $. A simple computation gives
\beeq
[J_0^5(x)\, ,\, \Psi (y)]\vert_{x^0=y^0} = - \frac{\sigma - \tau}{  \sigma +
\tau}\, \delta (x^1-y^1)\gamma^5 \Psi (y) \, .
\label{qc}
\end{equation}
Because of eq. (\ref{coup}), for $g\not= 0$ one has
\beeq
\sigma - \tau \not= \sigma + \tau \, ,
\end{equation}
which shows the presence of a quantum correction in the right hand side of
(\ref{qc}) with respect to (\ref{poi}). This is a known unavoidable feature
\cite{Swi}, following
from (a-b).

Finally, let us consider condition (d). From the results of Sect. 3.1
we deduce that it
is satisfied, if the parameters $\mu_{{}_Z}$ of $\T(\sigma , \tau ;
\mu_{{}_L}, \mu_{{}_R})$
are fixed according to
\beeq
\mu_{{}_R} = \frac{1}{ \sigma + \tau }\mu + \frac{1}{ \sigma - \tau }\mu_{{}_5}
\, , \qquad
\mu_{{}_L} = \frac{1}{ \sigma + \tau }\mu - \frac{1}{ \sigma - \tau }\mu_{{}_5}
\, .
\label{y}
\end{equation}
Using eqs. (\ref{par}) and (\ref{coup}) one can express $(\sigma,\, \tau )$
in terms of
$(g,\, \vartheta )$. Two solutions are found:
\bea
&&\sigma = \frac{g + 2\vartheta}{ 2\sqrt {g + \vartheta }} \, ,\qquad
\tau = \frac{g}{  2\sqrt {g + \vartheta }}\, ,  \label{sol} \\
&&\sigma^\prime = - \frac{g + 2\vartheta }{ 2\sqrt {g + \vartheta }}\, ,
\qquad \tau^\prime = - \frac{g}{  2\sqrt {g + \vartheta }} \, .
\label{solp}
\ena
Since $(\sigma,\, \tau )\in \R^2$, the coupling constant $g$ and the
statistical parameter $\vartheta $ are constrained by
\beeq
\vartheta > - g  \, .
\end{equation}
Anyonic solutions of the Thirring model have been proposed and investigated 
recently also in \cite{IT}. 

The correlation functions of the Thirring field $\Psi $ follow directly from 
eq. (\ref{ac}) and coincide for both solutions (\ref{sol}) and (\ref{solp}). From 
eqs. (\ref{exj},\ref{cthd},\ref{y}-\ref{solp}) one can derive
the expectation value of the Thirring current $J_\nu (x)$ in the
Gibbs state:
\beeq
\langle J_0(x)\rangle_{\mu,\, \mu_{{}_5}}^\beta =
\frac{\mu}{  \pi (g + \vartheta )} \, , \qquad
\langle J_1(x)\rangle_{\mu,\, \mu_{{}_5}}^\beta =
- \frac{\mu_{{}_5}}{  \pi \vartheta } \, .
\label{gibbs}
\end{equation}
Comparing eq. (\ref{gibbs}) to the free case (\ref{exc1},\ref{exc2}), 
one observes nontrivial corrections induced by the interaction and 
the anyon statistic. Previous investigations 
(\cite{Yo}, \cite{SW}, \cite{AEGN}) of the charge density 
$\langle J_0(x)\rangle_{\mu,\, \mu_{{}_5}}^\beta$ in
the case $\vartheta = 1$, have given contradictory results. We confirm
the result of \cite{SW} and \cite{AEGN} and extend it to the general class of 
anyon solutions with $\vartheta \not= 1$. In the functional integral framework 
of \cite{SW} and \cite{AEGN}, the $g$-dependence of the charge density stems from
topological contributions. It is worth stressing that our operator approach
reproduces the same $g$-dependence in a flat 1+1 dimensional Minkowski
space-time. 

To our  knowledge, the appearance of a persistent current
$\langle J_1(x)\rangle_{\mu,\, \mu_{{}_5}}^\beta $ in the finite temperature
Thirring model represents a novel feature. We would like to recall in 
this respect that persistent currents of quantum origin have been experimentally 
observed (\cite{LDDB}, \cite{MCB}) in mesoscopic rings placed in an external 
magnetic field. Such fields are absent in the two-dimensional world, but 
chiral symmetry, combined with duality still allow for a non-vanishing 
$\langle J_1(x)\rangle_{\mu,\, \mu_{{}_5}}^\beta $. Notice that the 
persistent current (\ref{gibbs}) grows for small values of $\vartheta $, 
diverging in the limit $\vartheta \to 0$. 

Concerning the energy-momentum tensor $T_{\mu \nu}(x)$ of the model,
by means of eq. (\ref{ta}) we find that
\bea
&&  T_{00}(x)  =    T_{11}(x) \equiv \Theta_{{}_L}(x^+)+\Theta_{{}_R}(x^-)
\, , \nb \\
&&T_{01}(x) =   T_{10}(x) \equiv \Theta_{{}_L}(x^+)
-\Theta_{{}_R}(x^-)\, ,
\ena
generate on the Thirring field $\Psi $ the right transformations
\beeq
[T_{0\nu}(x_1) \, , \, \Psi (x_2) ]\vert_{x_1^0 = x_2^0}  = - i \der_\nu
\Psi (x_2)\, \delta (x^1_{12}) \, .
\end{equation}
Employing eqs. (\ref{ex},\ref{tlim},\ref{y},\ref{gibbs}) one derives the
following energy and momentum densities:
\bea
&&\langle  T_{00}(x) \rangle_{\mu,\, \mu_{{}_5}}^\beta
= \frac{\pi}{6\beta^2} + \frac{\mu_{{}_L}^2 + \mu_{{}_R}^2}{ 4\pi} = \nb \\
&& \frac{\pi}{ 6\beta^2} + \frac{\pi}{ 2} (g+\vartheta)
\left[ \left (
\langle J_0(x)\rangle_{\mu,\, \mu_{{}_5}}^\beta \right )^2 +
\left (\langle J_1(x)\rangle_{\mu,\, \mu_{{}_5}}^\beta \right )^2 \right]
\, , \label{steq} \\
&&\langle  T_{01}(x) \rangle_{\mu,\, \mu_{{}_5}}^\beta
= \frac{\mu_{{}_L}^2 - \mu_{{}_R}^2}{ 4\pi} =
\pi (g+\vartheta) \langle J_0(x)\rangle_{\mu,\, \mu_{{}_5}}^\beta
\, \langle J_1(x)\rangle_{\mu,\, \mu_{{}_5}}^\beta
\, .
\ena
When $\langle J_1(x)\rangle_{\mu,\, \mu_{{}_5}}^\beta=0$, the energy
density coincides with the pressure and (\ref{steq}) gives rise to
an equation of state, relating temperature, pressure and density.

\section{\bf Conclusions}

In the present paper we have developed an operator formalism for bosonization at finite 
tempera\-ture and density, which applies not only to fermions, but 
covers the case of generalized statistics as well. The approach is systematic, 
works directly in infinite volume and produces an explicit solution. 
Starting from the conventional chiral field algebras $\{\A_{{}_Z}\, :\, Z=R,\, L\}$, 
naturally associated with the free massless scalar field
$\varphi$ and its dual $\widetilde{\varphi}$, 
we have constructed a set of fields $A(x,\xi)$, parametrized by
$\xi = (\sigma, \, \tau) \in \R^2$. For generic values of $\xi$,
$A(x,\xi)$ exhibits Abelian braid statistics. The basic tool in 
our framework is the thermal representation 
${\bf {\cal T}}_{{}_L}(\mu_{{}_L}) \otimes {\bf {\cal T}}_{{}_R}(\mu_{{}_R})$ of 
${ \bf {\cal A}}_{{}_L} \otimes { \bf {\cal A}}_{{}_R}$, characterized by inverse
temperature $\beta $ and chemical potentials $\mu_{{}_L}$ and $\mu_{{}_R}$. 
We derived explicitly all correlation functions of the field $A(x,\xi)$ 
in the representation $\T_{{}_L}(\mu_{{}_L}) \otimes \T_{{}_R}(\mu_{{}_R})$, 
verifying the KMS condition. 
The $n$-point function represents a generalization of the Jastrow-Laughlin 
wave function. The exact momentum distributions of the left- and right-moving 
modes, following from the 2-point function, 
reveal in certain range of the variables $(\sigma,\, \tau )$ a clear 
condensation-like behavior. The parameters $\mu_{{}_L}$ and $\mu_{{}_R}$ are related 
to special shift authomorphisms of ${\cal A}_{{}_L}$ and ${\cal A}_{{}_R}$. 
Suitable combinations of $\mu_{{}_L}$ and $\mu_{{}_R}$ define the 
physical chemical potentials $\mu $ and $\mu_{{}_5}$, which are proportional 
to the vector- and chiral-charge densities. By duality, $\mu_{{}_5}\not= 0$ implies  
a non-vanishing persistent current. 

The characteristic features of our finite temperature and density 
bosonization procedure have been illustrated on the example of the 
massless Thirring model. We established a general class of anyonic solutions, 
providing an unambiguous derivation of the charge density and demonstrating 
the presence of a persistent current. 

The general framework proposed in the paper is mathematically self-consistent 
and is interesting not only from a purely theoretical point of view. It 
is widely accepted by now, that the excitations 
described by the anyon field $A(x,\xi)$ are relevant for a class of low
dimensional condensed matter systems in which the one-dimensional Luttinger 
liquid plays an important role. It is claimed for instance that the edge currents 
in the fractional quantum Hall effect can be described by a Luttinger liquid
\cite{We}. Anderson's \cite{An} proposal for explaining high temperature superconductivity
is based on the two-dimensional Hubbard model, whose ground state and low-energy 
excitations are also interpreted in terms of a Luttinger liquid. In this context, 
the phenomenon of anyon condensation and the appearance of a persistent current, 
discovered in this paper, deserve further attention. For application to quantum 
impurity problems, it will be interesting to extend to finite temperature 
and density the bosonization procedure on the half line, developed in \cite{LiMin}. 
We hope to report on this subject in the near future.

\pagebreak

\vskip 3 truecm
\appendix
\centerline {\large \bf Appendix}
\vskip 1truecm
In this appendix we prove the determinant formulae (\ref{trick},\ref{trick1}).
It is obvious that
\beeq
{\rm det}\, \frac{1}{ (x_i-y_j)} =
\frac{P(x,y)}{ \prod_{{}_{i,j=1}}^n (x_i-y_j ) }
\, \, ,
\label{a1}
\end{equation}
where $P$ is a polynomial of degree $n^2 -n$, $n$ being the order of the
determinant. Because of translation invariance, $P$ depends only on
the differences $x_{ij}$, $y_{ij}$ and ${x_i-y_j}$. Moreover, if $x_i=x_j$ the
determinant has two identical rows and vanishes, so $P$ must be
divisible by $x_{ij}$;
in the same way if $y_i=y_j$
the determinant has two identical columns, so
$P$ must be divisible by $y_{ij}$. We can write then
\beeq
P(x,y)= C\, {\prod_{{}_{\stackrel{i,j=1}{ i<j}}}^n (x_{ij})\, \, 
\prod_{{}_{\stackrel{i,j=1}{i>j}}}^n(y_{ij})}
\label{a2}
\end{equation}
where $C$ is a constant, potentially depending on
$n$. Considering the particular case
$x_i= y_i + \epsilon_i $ and expanding in $\epsilon_i$ one immediately
checks that $C=1$. Formula (\ref{trick1}) is a direct consequence of
eqs. (\ref{a1},\ref{a2}).

Concerning the determinant formula (\ref{trick}), put
$z_i= e^{x_i} \, , \quad w_i= e^{y_i} \, . $ We have
\beeq
\det \frac{ 1}{ \sinh(x_i-y_j)}=
\det \frac{ 2 z_i w_j}{  z_i^2 - w_j^2}=
\frac{\prod_{{}_{i=1}}^n (2 z_i w_i)
\prod_{{}_{\stackrel{i,j=1}{ i<j}}}^n (z^2_i-z^2_j)\, \, \prod_{{}_{\stackrel{i,j=1}{i>j}}}^n(w^2_i -w^2_j)}{\prod_{{}_{i,j=1}}^n (z^2_i-w^2_j) } \, \, ,
\end{equation}
where the previous determinant formula has been used.
Now
\bea
&&z^2_i-z^2_j = 2 z_i z_j \sinh(x_{ij}) \, , \qquad
w^2_i-w^2_j = 2 w_i w_j \sinh(y_{ij}) \, , \nb \\
&& z^2_i-w^2_j = 2 z_i w_j \sinh(x_i-y_j) \, ,
\ena
and we get
\beeq
\frac{\prod_{{}_{i=1}}^n (2 z_i w_i)
\prod_{{}_{\stackrel{i,j=1}{i<j}}}^n (2 z_i z_j)
\, \, \prod_{{}_{\stackrel{i,j=1}{i>j}}}^n (2 w_i w_j)}{
 \prod_{{}_{i,j=1}}^n (2 z_i w_j)} \qquad
\frac{\prod_{{}_{\stackrel{i,j=1}{i<j}}}^n \sinh(x_{ij})
\, \, \prod_{{}_{\stackrel{i,j=1}{ i>j}}}^n \sinh (y_{ij} )}{
 \prod_{{}_{i,j=1}}^n
\sinh(x_i-y_j)} \, .
\end{equation}
A trivial calculation shows that the first fraction equals $1$
identically, so we finally obtain
\beeq
\det \frac{ 1}{ \sinh(x_i-y_j)}=
\frac{\prod_{{}_{\stackrel{i,j=1}{ i<j}}}^n \sinh(x_{ij})
\, \, \prod_{{}_{\stackrel{i,j=1}{ i>j}}}^n \sinh (y_{ij} )}
{\prod_{{}_{i,j=1}}^n \sinh(x_i-y_j)} \, .
\end{equation}
Eq. (\ref{trick}) is a straightforward consequence of this identity.

\end{document}